\documentclass[journal]{IEEEtran}

\usepackage{times}
\usepackage{epsfig}
\usepackage{multirow}
\usepackage{changepage}
\usepackage{pifont}
\usepackage{amssymb}
\usepackage{pdflscape}
\usepackage{supertabular}
\usepackage{xcolor}
\usepackage{pgfplots}
\usepackage{float}
\usepackage{subfig}
\usepackage{amsmath}
\usetikzlibrary{shadows}
\usepackage{colortbl}
\usepackage{tikz}
\usepackage{amsthm} 
\usepackage{hyperref}
\usepackage{soul} 
\usepackage{graphicx}
\usetikzlibrary{arrows,automata,backgrounds,chains,fit,calc,spy}

\newenvironment{customlegend}[1][]{%
    \begingroup
    \csname pgfplots@init@cleared@structures\endcsname
    \pgfplotsset{#1}%
}{%
    \csname pgfplots@createlegend\endcsname
    \endgroup
}%
\graphicspath{{images/}}

\def\addlegendimage{\csname pgfplots@addlegendimage\endcsname}

\theoremstyle{definition}
\newtheorem*{defn}{Definition} %

\begin{document}
%
\title{Cloud Elasticity Using \\Probabilistic Model Checking}
%
%
%

\author{\IEEEauthorblockN{Athanasios Naskos\IEEEauthorrefmark{1}, Emmanouela Stachtiari\IEEEauthorrefmark{1}, Anastasios Gounaris\IEEEauthorrefmark{1}, Panagiotis Katsaros\IEEEauthorrefmark{1},\\
Dimitrios Tsoumakos\IEEEauthorrefmark{3}, Ioannis Konstantinou\IEEEauthorrefmark{2} and Spyros Sioutas\IEEEauthorrefmark{3}}\\
\IEEEauthorblockA{\IEEEauthorrefmark{1}Aristotle University of Thessaloniki, Greece, Email: \{anaskos,emmastac,gounaria,katsaros\}@csd.auth.gr}\\
\IEEEauthorblockA{\IEEEauthorrefmark{2}National Technical University of Athens, Greece, Email: ikons@cslab.ece.ntua.gr}\\
\IEEEauthorblockA{\IEEEauthorrefmark{3}Ionian University, Greece, Email: \{dtsouma,sioutas\}@ionio.gr}
}

\maketitle

\begin{abstract}
Cloud computing has become the leading paradigm for deploying large-scale infrastructures and  running big data applications, due to its capacity of achieving economies of scale. In this work, we focus on one of the most prominent advantages of cloud computing, namely the on-demand resource provisioning, which is commonly referred to as elasticity. Although a lot of effort has been invested in developing systems and mechanisms that enable elasticity, the elasticity decision policies tend to be designed without guaranteeing or quantifying the quality of their operation. This work aims to make the development of elasticity policies more formalized and dependable. We make two distinct contributions. First, we propose an extensible approach to enforcing elasticity through the dynamic instantiation and online quantitative verification of Markov Decision Processes (MDP) using probabilistic model checking.
Second, we propose concrete elasticity models and related elasticity policies. We evaluate our decision policies using both real and synthetic datasets in clusters of NoSQL databases. According to the experimental results,  our approach improves upon the state-of-the-art in significantly increasing user-defined utility values and decreasing user-defined threshold violations.
\end{abstract}


%
\IEEEpeerreviewmaketitle

\section{Introduction}
\label{sec.intro}
\IEEEPARstart{C}{loud} computing has arisen as one of the most attractive alternatives for providing computational infrastructures for high-demand applications.
The quick prevalence of clouds is fueled by their capacity of achieving economies of scale. One of the main advantages of cloud computing is that it renders the procurement of
expensive computing resources unnecessary, thus lifting the burden  of high upfront investments in proprietary platforms from system developers and owners. This characteristic is complemented by the capacity for on-demand resource provisioning based on the actual current requirements; this feature is commonly referred to as elasticity, and it is the main focus of this work.

Elasticity plays an important role in cloud-based provisioning of resources for big data applications. This is because elasticity is the main mechanism through which cloud resource
provisioning methods are capable of scaling and performing well under highly unpredictable conditions. Both these characteristics, i.e., scalability and adaptations to volatile conditions, are particularly important when the data volume to be processed can be extremely large.

Cloud resource elasticity may be applied in different forms and can refer to the size, the location or the number of virtual machines (VMs) employed. Examples of these three elasticity types are the allocation of more memory to a VM, moving a VM to a less loaded physical machine and increasing the number of VMs of an application cluster, respectively. Interestingly, resource elasticity does not necessarily comprise the notion of automation. In non-automated settings, users should know in advance the resource needs of their applications (e.g., number of VMs) and they should either schedule elasticity actions or continuously monitor the state of their applications in order to decide whether a change in the amount of resources is needed.
Here, we exclusively focus on automated elasticity approaches,
and we especially target elasticity in the form of horizontally scaling the number of application VMs. Increasing or decreasing the number of VMs is a key element in adapting to dynamically changing volumes of user requests, e.g., as typically occurs in cloud databases, which is the scenario we used in our evaluation.

We adopt the standard elasticity definition proposed in \cite{herbst2013elasticity_definition}:
\begin{defn}Elasticity is the degree to which a system is able to adapt to workload changes by provisioning and de-provisioning resources in an autonomic manner, such that at each point in time the available resources match the current demand as closely as possible.\end{defn}

Recently, there have been numerous proposals for elasticity, which differ in several dimensions including the form of the elasticity they support, the underlying objectives driving the elasticity actions and the decision making policy (e.g., reactive or proactive), such as \cite{tsoumakos2013automated,gandhi_autoscale:_2012,shen_cloudscale:_2011,gong_press:_2010,trushkowsky_scads_2011,Bairavasundaram:2012Dynamite}. However, elasticity proposals tend to not be accompanied by correctness guarantees. The main aim of our proposal is to make a decisive step towards more dependable and formalized elasticity decision policies. By more dependable we mean that we decide elasticity actions according to the results of continuous verification of key elasticity aspects, including the resulting system utility.
At a higher level, we view the elasticity problem as a specific instance of autonomic computing \cite{KC03}, for which the need for coupling continuous verification when responding to environmental changes has already been identified \cite{CGK+12}. Formal verification applies mathematical reasoning in order to provide correctness guarantees; to this end, in this work, we adopt a successful verification method, namely (probabilistic) model checking \cite{FKNP11}.

In brief, our approach consists of two steps. First, we present expressive models of elasticity actions and second, we leverage them for devising concrete policies that can take elasticity decisions. The mathematical modeling framework we adopt is Markov Decision Processes (MDPs), because MDPs can capture both the non-deterministic and probabilistic aspects of the problem. Non-determinism is due to the applicability of several possible elasticity actions, whereas the probabilistic behavior allows us to take into account the effects of the unpredictable environment's evolution.
In addition, we use the PRISM probabilistic model checker tool \cite{kwiatkowska2009Prism}, because it both supports the specification of MDP system models at a high-level
and comes with a property specification language called PCTL \cite{FKNP11}, for specifying probabilistic reachability and reward-based properties, which are amenable to model checking.
We introduce properties that, on the model level, yield optimal decisions for system reconfigurations aiming to maximize the system utility. We show how our decision making policy can be incorporated into existing systems. Finally, we discuss the Amazon's EC2 manager and the novel Tiramola system\footnote{Best-paper award in 2013 IEEE/ACM International Conference on Cluster, Cloud and Grid Computing.}, which supports elastic scaling of NoSQL databases \cite{tsoumakos2013automated}. In summary, the main contributions are:
\begin{enumerate}

\item We present a concrete approach to employing continuous online quantitative verification for taking elasticity decisions, with a view to making them more dependable. Our approach is well-founded and is based on extensible, automatically generated and dynamically instantiated MDP models.

\item We present modelling variations and related elasticity decision policies that aim to maximize user-defined utility functions; moreover, our decisions are subject to quantitative analysis.

\item We conduct thorough experiments using both real and synthetic data referring to an elastic cluster of NoSQL databases, which is the typical storage choice for large-scale applications. The evaluation results show that we can significantly increase user-defined utility values, which penalise over-provisioning i.e. providing more VMs than necessary, and decrease the frequency of user-defined threshold violations.

\end{enumerate}

The remainder of this paper is structured as follows. In Section \ref{sec:model}, we present our whole approach to elasticity decision making. We introduce three underlying MDP models and the elasticity policies that are built on top of their runtime instantiations. We also explain how the PRISM tool can be used for this purpose and provide examples of quantitative verification. Next, we discuss how our approach can be incorporated into existing systems. We evaluate our decision making solutions in Section \ref{sec:exps}, we refer to the related work in Section \ref{sec:rw} and we present future extensions of our work and conclusions in Section \ref{sec:conclusions}. In the Appendix, more details are provided about the Prism model structure and experimental configuration.

\section{The probabilistic model checking-based approach}
\label{sec:model}

Probabilistic model checking is a formal verification technique for the modelling and analysis of stochastic systems \cite{kwiatkowska2009Prism}. In our work, probabilistic models are used in the decision making process, to describe, drive and analyse cloud resource elasticity. By utilizing probabilistic models, we are able to capture the uncertain behaviour of systems elasticity. In order to additionally capture non-determinism we resort to MDP models, which form the basis of our approach. Similarly to our implementation, there are numerous other approaches where MDPs are used to handle both runtime and offline decision making, see Section \ref{sec:rw}. On top of our MDP models, we build policies for elasticity decisions with the help of the PRISM probabilistic model checker \cite{kwiatkowska2009Prism}. The form of elasticity that we consider is the resizing of a cluster, i.e., dynamically modifying the number of VMs with a view to optimizing a utility function. While our main objective is to render elasticity decision policies more dependable, our principled approach is capable of yielding higher utility, as will be shown in the evaluation section.

In cluster resizing, the elasticity decisions are typically bounded according to user-specified limits, so that not too many VMs are added or removed in a single step. This constraint corresponds to a technical requirement to be met in elastic systems \cite{herbst2013elasticity_definition}. The minimum and maximum  number of possible active VMs can be set based on preliminary analysis. Finally, we assume that the elasticity objectives are appropriately captured by a utility function, which is capable of assessing the quality of a specific cluster configuration. For an example of such a function, consider a user evaluating financial cost and gains for his application; cost is tied to the amount of resources commissioned from the underlying IaaS. Gains can relate to throughput, which is affected by the total number of active VMs.
In the remainder of this section, we deal with the formulation of MDPs, three model flavours, on top of which we devise concrete elasticity policies, examples of utility functions and quantitative analysis.

\subsection{Background}

MDPs serve as a powerful tool for the elasticity decision making process, as they provide a mathematical framework for modelling decision making in situations, where outcomes are partly random and partly under the control of a decision maker \cite{puterman1994MDP}. This condition fits well into our application domain, where we need a) to take decisions among multiple options, i.e,  adding or removing or maintaining the number of VMs and b) to maximize a utility function that quantifies the value of each system state, which is constantly evolving and hard, if not impossible, to be accurately predicted.

A Markov decision process (MDP) is a tuple \( M = (S,s_{init},Act,P_{sas'},L,R)\),  where
\begin{itemize}
\item $S = \{s_0,...,s_n\}$ is a finite set of states;
\item $s_{init}$ the initial state;
\item $Act= \{a_0,...,a_m\}$ is a finite set of actions;
\item $P_{sas'}=Pr\{s^{t+1}=s'|s^t=s,  a \in Act\}$ is a transition probability from state $s$ at step $t$ to state $s'$ at the next step due to action $a$;
\item $L$ a finite set of state labels; and
\item $R = (r_s,r_a)$ is a pair, the elements of which denote the rewards assigned to states and transitions between states, respectively. Formally, the state rewards are defined as
$r_s : S \rightarrow \operatorname*{\mathbb{R}}_{\geq 0}$ and the action rewards as $r_a : S \times S \times Act \rightarrow \operatorname*{\mathbb{R}}_{\geq 0}$. The total reward is given by the aggregation of the state and action rewards.
\end{itemize}


\begin{figure}[tb!]
\centering
\begin{tikzpicture}[->,>=stealth',shorten >=1pt,auto,inner sep=0pt,scale=0.9, thin ,initial text=current,initial where=right,every state/.style={draw=black,thin,minimum width=.4cm,circular drop shadow,fill=white}]

	\node[state,label={[brown,shift={(0.2,0)}]0:[5vms]}] (s5) {$s_5$};

	\node[initial,state,label={[brown,shift={(0.2,0.2)}]30:[4vms]}] (s4) at ($(s5.center)+(2.5,2.5cm)$) {$s_4$};	
	
	\node[state,label={[brown,shift={(-0.1,0)}]180:[3vms]}] (s3) at ($(s5.center)+(-2.5,2.5cm)$) {$s_3$};
	
	\node[state,label={[brown,shift={(-0.3,0.5)}]230:[6vms]}] (s6) at ($(s5.center)+(2.5,-2.5cm)$) {$s_6$};
	
	\node[state,label={[brown,shift={(-0.1,0)}]180:[7vms]}] (s7) at ($(s5.center)+(-2.5,-2.5cm)$) {$s_7$};
	
	\path
		(s3) edge[loop above, distance=10mm] node [pos=0.5, above=.1]{$no\_op:1$} (s3)
		(s4) edge[loop above, distance=10mm] node [pos=0.5, above=.1]{$no\_op:1$} (s4)
		(s5) edge[loop left, distance=10mm,sloped,anchor=south] node [pos=0.5, above=.1]{$no\_op:1$} (s5)
		(s6) edge[loop below, distance=10mm] node [pos=0.5, below=.1]{$no\_op:1$} (s6)
		(s7) edge[loop above, distance=10mm,sloped,anchor=south] node [pos=0.5, above=.1]{$no\_op:1$} (s7)
		
		(s3) edge[bend left=20,sloped,anchor=south, blue] node [pos=0.5, above=.1]{$add_1:0.5$} (s4)
		(s3) edge[bend left=20,sloped,anchor=south, blue] node [pos=0.5, below=.1]{$add_2:0.5$} (s5)
		
		(s4) edge[bend left=20,sloped,anchor=south, blue] node [pos=0.5, above=.1]{$add_1:0.5$} (s5)
		(s4) edge[bend left=20,sloped,anchor=south, blue] node [pos=0.5, above=.1]{$add_2:0.5$} (s6)
		
		(s5) edge[bend left=20,sloped,anchor=south, blue] node [pos=0.5, above=.1]{$add_1:0.5$} (s6)
		(s5) edge[bend left=20,sloped,anchor=south, blue] node [pos=0.5, above=.1]{$add_2:0.5$} (s7)
		
		(s6) edge[bend left=20,sloped,anchor=south, blue] node [pos=0.5, above=.1]{$add_1:1$} (s7)

		(s4) edge[bend left=20,sloped,anchor=south, red] node [pos=0.5, above=.1]{$rem_1:1$} (s3)

		(s5) edge[bend left=20,sloped,anchor=south, red] node [pos=0.3, above=.1]{$rem_1:1$} (s4)

		(s6) edge[bend left=20,sloped,anchor=south, red] node [pos=0.5, above=.1]{$rem_1:1$} (s5)

		(s7) edge[bend left=20,sloped,anchor=south, red] node [pos=0.5, above=.1]{$rem_1:1$} (s6)
	;
	\begin{customlegend}[
		legend entries={ 
			$add$,
			$remove$,
			$no\_op$
		},
	legend style={at={(5.7,-2))},font=\tiny}] 
	    \addlegendimage{-,blue,thick}
	    \addlegendimage{-,red,thick}
	    \addlegendimage{-,black,thick}
	\end{customlegend}
\end{tikzpicture}
\caption{MDP model overview.}
\label{fig:model_original}
\end{figure}
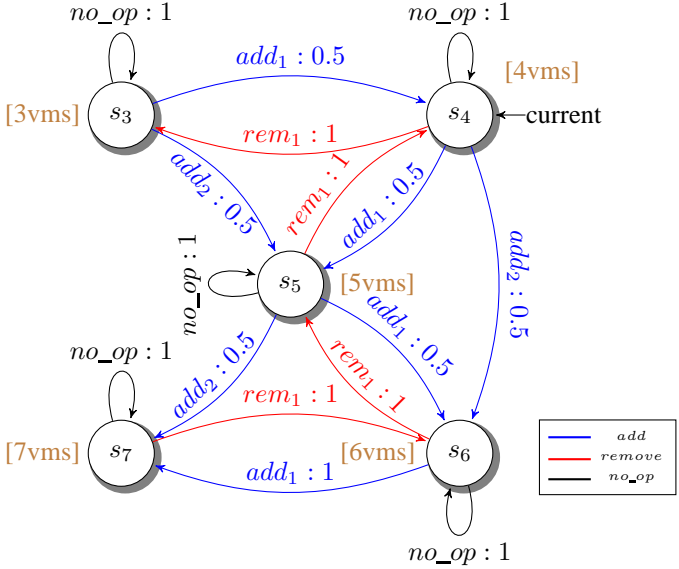

\subsection{Elasticity Models}
In this section we present a simplified view of our modelling approach. The actual model implementation details are presented in Appendix.

In our model, each state corresponds to a different cluster size, where the size equals to the number of active VMs, ${vms\_num}$. That size forms the state label. If $|$VM$_{min}|$ (resp. $|$VM$_{max}|$) is the minimum (resp. maximum) valid size of our infrastructure, then there are at most $|$\textit{VM$_{max}$}$|$-$|$\textit{VM$_{min}$}$|$+1 states, i.e., the model is adequately small to be analyzed efficiently \cite{kwiatkowska2009Prism}.
For readability reasons, we denote a state as  $s_{[vms\_num]}$.  There are three general types of model actions, which give rise to non-deterministic behaviour: 1) $add$ for VM additions, 2) $rem$ for removals, and 3) $no\_op$ for no operation. Every elasticity decision, that is every possible scale-up or scale-down in the number of the active VMs, is represented as a transition between two states of the model. In addition, for each action type, there may be multiple valid transitions; for example, we can decide to $add$ 2, 3 or more VMs. All the transitions are mapped to a probability. As the MDP dictates, the probabilities of a specific type of action from the same state are summed to 1. By default, the transitions of the same action type are equally probable in order not to bias the system.

The MDP associates a reward value to each state and action. State and action rewards are calculated based on user-specified utility functions.
When the model is verified at runtime, the utility at state $s_{[vms\_num]}$ essentially describes the expected behaviour of the system when there are $vms\_num$ active VMs.
We use the utility functions in order to derive the reward of each state, and we defer the presentation of example utility functions later in this section.

\begin{figure}[tb!]
\begin{center}
\begin{tikzpicture}
	\node[drop shadow] (table) [inner sep=0pt] {
		\begin{tabular}{ >{\centering\arraybackslash} p{7.25cm}}
			\rowcolor[HTML]{262686}
  			$\boldsymbol{\textcolor{white}{M = (S,s_{init},Act,P_{sas'},L,R)}}$ \\[1ex]
  			\hline
  			\rowcolor[HTML]{EAEAF3}
  			\begin{itemize}
				\item $S = \{s_3,s_4,s_5,s_6,s_7\}$
				\item $s_{init} = s_4$
				\item $Act = \{add,rem,no\_op\}$
				\item $P_{s,add,s'}=$
				\begin{tabular}{|c c c c c|}
				0&0.5&0.5&0&0 \\
				0&0&0.5&0.5&0 \\
				0&0&0&0.5&0.5 \\
				0&0&0&0&1 \\
				0&0&0&0&0 \\
				\end{tabular}
				
                \item $P_{s,rem,s'}=$
				\begin{tabular}{|c c c c c|}
				0&0&0&0&0 \\
				1&0&0&0&0 \\
				0&1&0&0&0 \\
				0&0&1&0&0 \\
				0&0&0&1&0 \\
				\end{tabular}
				
                \item $P_{s,no\_op,s'}= I_5=diag(1,1,1,1,1)$
				\item $L=\{3vms,4vms,5vms,6vms,7vms\}$
				\item $r_s = utility\_function(s)$, $r_a=0$
				\end{itemize}
		\end{tabular}
	};
\end{tikzpicture}
\end{center}
\caption{MDP instantiation for the model in Figure \ref{fig:model_original}.}
\label{fig:mdp_tuple}
\end{figure}
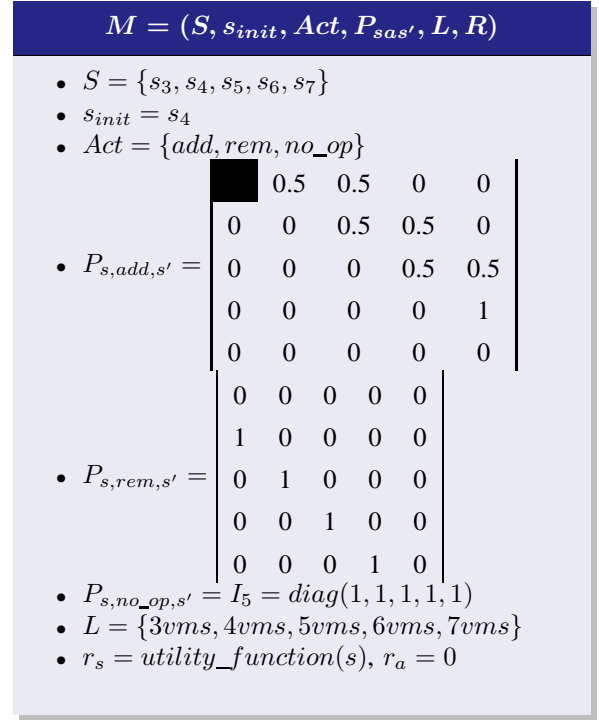

Figure \ref{fig:model_original} illustrates a simplified instance of the MDP model, where the states represent the number of active VMs while the edges represent the possible actions: 1) $add_{new\_vms\_num}$ (blue arrow), 2) $rem_{removed\_vms\_num}$ (red arrow), and 3) $no\_op$ (black arrow). In this example, the maximum number of VMs allowed to be added or removed in every step is 2 and 1, respectively, while the current state is $s_4$. The action type combined with its probability is labelled on top of every transition (${[add_x/rem_x/no\_op]}:P_{[add_x/rem_x/no\_op]}$) while the label of every state (brown color) is beside it. The MDP instantiation of that model is presented in Figure \ref{fig:mdp_tuple}.

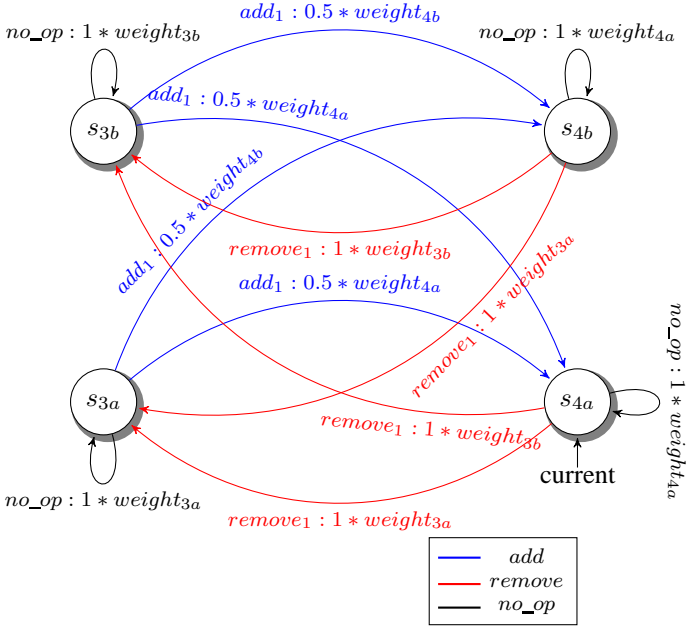
\begin{figure}[tb!]
\centering
\begin{tikzpicture}[->,>=stealth',shorten >=1pt,auto,inner sep=0pt, thin,scale=0.9 ,initial text=current,initial where=below,every state/.style={draw=black,thin,minimum width=.3cm,circular drop shadow,fill=white}]

	\node[initial,state] (s4cl0) at (3.5,0cm) {$s_{4a}$};	
	
	\node[state] (s3cl0) at (-3.5,0cm) {$s_{3a}$};
	
	\node[state] (s4cl1) at (3.5,4cm) {$s_{4b}$};
	
	\node[state] (s3cl1) at (-3.5,4cm) {$s_{3b}$};
	
	\path
		(s3cl0) edge[loop below,sloped,anchor=south, distance=10mm] node [pos=0.5, below=.1]{\footnotesize $no\_op:1*weight_{3a}$} (s3cl0)
		(s4cl0) edge[loop right,sloped,anchor=south, distance=10mm] node [pos=0.5, above=.1]{\footnotesize $no\_op:1*weight_{4a}$} (s4cl0)
		(s3cl1) edge[loop above, distance=10mm] node [pos=0.5, above=.1]{\footnotesize $no\_op:1*weight_{3b}$} (s3cl1)
		(s4cl1) edge[loop above, distance=10mm] node [pos=0.5, above=.1]{\footnotesize $no\_op:1*weight_{4a}$} (s4cl1)
		
		(s3cl0) edge[bend left=40,sloped,anchor=south, blue] node [pos=0.5, above=.1]{\footnotesize $add_1:0.5*weight_{4a}$} (s4cl0)
		(s3cl0) edge[bend left=40,sloped,anchor=south, blue] node [pos=0.3, above=.1]{\footnotesize $add_1:0.5*weight_{4b}$} (s4cl1)
		(s3cl1) edge[bend left=40,sloped,anchor=south, blue] node [pos=0.2, above=.1]{\footnotesize $add_1:0.5*weight_{4a}$} (s4cl0)
		(s3cl1) edge[bend left=40,sloped,anchor=south, blue] node [pos=0.5, above=.1]{\footnotesize $add_1:0.5*weight_{4b}$} (s4cl1)

		(s4cl0) edge[bend left=40,sloped,anchor=south, red] node [pos=0.5, below=.1]{\footnotesize $remove_1:1*weight_{3a}$} (s3cl0)
		(s4cl0) edge[bend left=40,sloped,anchor=south, red] node [pos=0.2, below=.1]{\footnotesize $remove_1:1*weight_{3b}$} (s3cl1)
		(s4cl1) edge[bend left=40,sloped,anchor=south, red] node [pos=0.3, below=.1]{\footnotesize $remove_1:1*weight_{3a}$} (s3cl0)
		(s4cl1) edge[bend left=40,sloped,anchor=south, red] node [pos=0.5, below=.1]{\footnotesize $remove_1:1*weight_{3b}$} (s3cl1)
	;

	\begin{customlegend}[
		legend entries={ 
			$add$,
			$remove$,
			$no\_op$
		},
	legend style={at={(3.5,-2))},font=\footnotesize}] 
	    \addlegendimage{-,blue, thick}
	    \addlegendimage{-,red, thick}
	    \addlegendimage{-,black,thick}
	\end{customlegend}
\end{tikzpicture}
\caption{Model extension considering multiple states per number of VMs.} \label{fig:multi_cluster}
\end{figure}

The model of Figure \ref{fig:model_original} may be deemed as a simplified version, which can be more elaborated with a view to considering additional factors. More specifically, it implies that each state is associated with a single reward and assumes that the reward function depends on system latency and throughput. However, for the same amount of VMs, the latency and throughput may vary significantly, due to external factors. This leads to an undesirable situation, where the state reward does not reflect the actual system behaviour well, and there is a high probability that the system may end up in a real state that significantly diverges from the one originally expected. To ameliorate this, we can increase the number of model states that correspond to a specific number of VMs, so that, each state corresponds to a distinct expected behaviour for that amount.
Moreover, we also extend the model so that it explicitly covers the probabilities of encountering each of the new states.
Figure \ref{fig:multi_cluster} shows the upper part of the model in Figure \ref{fig:model_original}, where there are two states for each size, $a$ and $b$, and the transitions are enriched with the probabilities according to state weights. The state weights are equal to their probability. Thus, when a MDP solver examines possible actions to maximize the total reward, it can better capture the fact that the behaviour of the system is non-deterministic and unpredictable. Nevertheless, the higher expressivity of the model comes at the expense of larger state space size, compared to the simpler model. However, according to Section \ref{sec:exps_overhead}, the additional overhead is negligible. More details about the number of states per VM number are given when we discuss state reward specification, where the notion of clusters is added to our model.

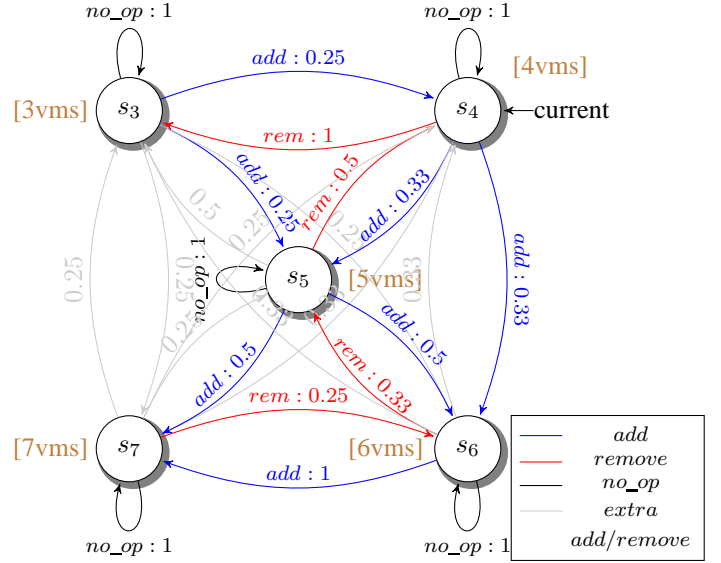
\begin{figure}[tb!]
\centering
\begin{tikzpicture}[->,>=stealth',shorten >=1pt,auto,inner sep=0pt, thin,scale=0.9 ,initial text=current,initial where=right,every state/.style={draw=black,thin,minimum width=.4cm,circular drop shadow,fill=white}]

	\node[state,label={[brown,shift={(0.2,0)}]0:[5vms]}] (s5) {$s_5$};

	\node[initial,state,label={[brown,shift={(0.2,0.2)}]30:[4vms]}] (s4) at ($(s5.center)+(2.5,2.5cm)$) {$s_4$};	
	
	\node[state,label={[brown,shift={(-0.1,0)}]180:[3vms]}] (s3) at ($(s5.center)+(-2.5,2.5cm)$) {$s_3$};
	
	\node[state,label={[brown,shift={(-0.3,0.5)}]230:[6vms]}] (s6) at ($(s5.center)+(2.5,-2.5cm)$) {$s_6$};
	
	\node[state,label={[brown,shift={(-0.1,0)}]180:[7vms]}] (s7) at ($(s5.center)+(-2.5,-2.5cm)$) {$s_7$};
	
	\path
		(s3) edge[loop above, distance=10mm] node [pos=0.5, above=.1]{\footnotesize $no\_op:1$} (s3)
		(s4) edge[loop above, distance=10mm] node [pos=0.5, above=.1]{\footnotesize $no\_op:1$} (s4)
		(s5) edge[loop left, distance=10mm,sloped,anchor=south] node [pos=0.5, above=.1]{\footnotesize $no\_op:1$} (s5)
		(s6) edge[loop below, distance=10mm] node [pos=0.5, below=.1]{\footnotesize $no\_op:1$} (s6)
		(s7) edge[loop below, distance=10mm,sloped,anchor=south] node [pos=0.5, below=.1]{\footnotesize $no\_op:1$} (s7)
		
		(s3) edge[bend left=20,sloped,anchor=south, blue] node [pos=0.5, above=.1]{\footnotesize $add:0.25$} (s4)
		(s3) edge[bend left=20,sloped,anchor=south, blue] node [pos=0.7, above=.1]{\footnotesize $add:0.25$} (s5)
		(s3) edge[bend left=20,sloped,anchor=south, gray!40] node [pos=0.5, above=.1]{$0.25$} (s6)
		(s3) edge[bend left=20,sloped,anchor=south, gray!40] node [pos=0.5, above=.1]{$0.25$} (s7)
		
		(s4) edge[bend left=20,sloped,anchor=south, blue] node [pos=0.5, above=.1]{\footnotesize $add:0.33$} (s5)
		(s4) edge[bend left=20,sloped,anchor=south, blue] node [pos=0.5, above=.1]{\footnotesize $add:0.33$} (s6)
		(s4) edge[bend left=20,sloped,anchor=south, gray!40] node [pos=0.5, above=.1]{$0.33$} (s7)
		
		(s5) edge[bend left=20,sloped,anchor=south, blue] node [pos=0.5, above=.1]{\footnotesize $add:0.5$} (s6)
		(s5) edge[bend left=20,sloped,anchor=south, blue] node [pos=0.5, above=.1]{\footnotesize $add:0.5$} (s7)
		
		(s6) edge[bend left=20,sloped,anchor=south, blue] node [pos=0.5, above=.1]{\footnotesize $add:1$} (s7)

		(s4) edge[bend left=20,sloped,anchor=south, red] node [pos=0.5, above=.1]{\footnotesize $rem:1$} (s3)

		(s5) edge[bend left=20,sloped,anchor=south, red] node [pos=0.3, above=.1]{\footnotesize $rem:0.5$} (s4)
		(s5) edge[bend left=20,sloped,anchor=south, gray!40] node [pos=0.5, above=.1]{$0.5$} (s3)

		(s6) edge[bend left=20,sloped,anchor=south, red] node [pos=0.5, above=.1]{\footnotesize $rem:0.33$} (s5)
		(s6) edge[bend left=20,sloped,anchor=south, gray!40] node [pos=0.5, above=.1]{$0.33$} (s4)
		(s6) edge[bend left=20,sloped,anchor=south, gray!40] node [pos=0.5, above=.1]{$0.33$} (s3)

		(s7) edge[bend left=20,sloped,anchor=south, red] node [pos=0.5, above=.1]{\footnotesize $rem:0.25$} (s6)
		(s7) edge[bend left=20,sloped,anchor=south, gray!40] node [pos=0.5, above=.1]{$0.25$} (s5)
		(s7) edge[bend left=20,sloped,anchor=south, gray!40] node [pos=0.5, above=.1]{$0.25$} (s4)
		(s7) edge[bend left=20,sloped,anchor=south, gray!40] node [pos=0.5, above=.1]{$0.25$} (s3)
	;

	\begin{customlegend}[
		legend entries={ 
			$add$,
			$remove$,
			$no\_op$,
			$extra$,
            $add/remove$
		},
	legend style={at={(6,-2))},font=\footnotesize}] 
	    \addlegendimage{-,blue}
	    \addlegendimage{-,red}
	    \addlegendimage{-,black}
	    \addlegendimage{-,gray!40}
        \addlegendimage{,white}
	\end{customlegend}
\end{tikzpicture}
\caption{Model extension allowing arbitrary inter-state transitions.} \label{fig:looking_forward}
\end{figure}

A third model variation allows the transitions between states to disregard the limits on the maximum number of VMs allowed to add or remove in each step. Figure \ref{fig:looking_forward} illustrates an example of this enhanced type of model, which contains the additional transitions that are not possible to be immediately enforced (in light grey color).
The rationale behind that technique is the investigation of potential benefits in a more brute force manner as it does not take into consideration only the actual accessible states but all the states in the model. If the model solver indicates a state which is not currently accessible as the most beneficial, the actual action will be bounded according to the user-specified add/remove limits\footnote{The discussion in this section has regarded the models mostly at the conceptual level. In order to implement the models according to the PRISM's specifications, additional issues need to be considered, which are omitted because they do not contribute to the discussion.}.

\subsection{Policies for Elasticity Decisions}
The previous section discusses how we can model the elasticity actions, where here we present exact approaches to taking elasticity decisions. In general, we periodically monitor the incoming load and the system state; also, we periodically activate the decision policy, and we call such an activation an elasticity step. The monitoring frequency is typically higher than the decision making frequency. An elasticity step is further split in the following three sub-phases:

\begin{enumerate}

\item Dynamically instantiate a model according to the current incoming load and the log measurements; the current load influences the expected values of the utility function of each state, which in turn, specify the state rewards.

\item Verify the model online.

\item Take elasticity actions.

\end{enumerate}

The first subphase is the most important one. The model is dynamically instantiated so that, in each step, it can describe the expected behavior according to the current environmental conditions. In our implementation, those conditions are defined by the (external) incoming load  $\lambda$ of requests. We assume that the system that sets the elasticity decision policy keeps log measurements in order to be capable of evaluating the utility functions given the current value of $\lambda$.

\subsubsection{Indirect vs. Direct Solutions}

There are several options to analyze the MDPs. We distinguish between  indirect (based on reinforcement learning) and direct methods (based on dynamic programming).

Indirect methods are exemplified by the Tiramola approach, which relies on online training and convergence of action-value functions, which, in turn allows to attain optimal policies through greedy actions and a Q-learning-based reinforcement learning approach. Exact details are provided in \cite{tsoumakos2013automated}.

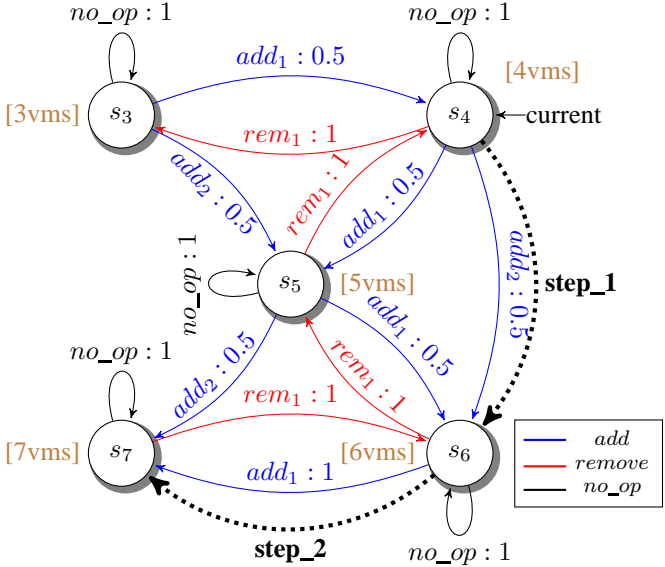
\begin{figure}[tb!]
\centering
\begin{tikzpicture}[->,>=stealth',shorten >=1pt,auto,inner sep=0pt,scale=0.9, thin ,initial text=current,initial where=right,every state/.style={draw=black,thin,minimum width=.4cm,circular drop shadow,fill=white}]

	\node[state,label={[brown,shift={(0.2,0)}]0:[5vms]}] (s5) {$s_5$};

	\node[initial,state,label={[brown,shift={(0.2,0.2)}]30:[4vms]}] (s4) at ($(s5.center)+(2.5,2.5cm)$) {$s_4$};	
	
	\node[state,label={[brown,shift={(-0.1,0)}]180:[3vms]}] (s3) at ($(s5.center)+(-2.5,2.5cm)$) {$s_3$};
	
	\node[state,label={[brown,shift={(-0.3,0.5)}]230:[6vms]}] (s6) at ($(s5.center)+(2.5,-2.5cm)$) {$s_6$};
	
	\node[state,label={[brown,shift={(-0.1,0)}]180:[7vms]}] (s7) at ($(s5.center)+(-2.5,-2.5cm)$) {$s_7$};
	
	\path
		(s3) edge[loop above, distance=10mm] node [pos=0.5, above=.1]{$no\_op:1$} (s3)
		(s4) edge[loop above, distance=10mm] node [pos=0.5, above=.1]{$no\_op:1$} (s4)
		(s5) edge[loop left, distance=10mm,sloped,anchor=south] node [pos=0.5, above=.1]{$no\_op:1$} (s5)
		(s6) edge[loop below, distance=10mm] node [pos=0.5, below=.1]{$no\_op:1$} (s6)
		(s7) edge[loop above, distance=10mm,sloped,anchor=south] node [pos=0.5, above=.1]{$no\_op:1$} (s7)
		
		(s3) edge[bend left=20,sloped,anchor=south, blue] node [pos=0.5, above=.1]{$add_1:0.5$} (s4)
		(s3) edge[bend left=20,sloped,anchor=south, blue] node [pos=0.5, below=.1]{$add_2:0.5$} (s5)
		
		(s4) edge[bend left=20,sloped,anchor=south, blue] node [pos=0.5, above=.1]{$add_1:0.5$} (s5)
		(s4) edge[bend left=20,sloped,anchor=south, blue] node [pos=0.5, above=.1]{$add_2:0.5$} (s6)
		(s4) edge[bend left=40,dotted,ultra thick,black] node [pos=0.5, right=0.1]{\textbf{step\_1}} (s6)
		
		(s5) edge[bend left=20,sloped,anchor=south, blue] node [pos=0.5, above=.1]{$add_1:0.5$} (s6)
		(s5) edge[bend left=20,sloped,anchor=south, blue] node [pos=0.5, above=.1]{$add_2:0.5$} (s7)
		
		(s6) edge[bend left=20,sloped,anchor=south, blue] node [pos=0.5, above=.1]{$add_1:1$} (s7)
		(s6) edge[bend left=40,dotted,ultra thick,black] node [pos=0.5, below=0.1]{\textbf{step\_2}} (s7)

		(s4) edge[bend left=20,sloped,anchor=south, red] node [pos=0.5, above=.1]{$rem_1:1$} (s3)

		(s5) edge[bend left=20,sloped,anchor=south, red] node [pos=0.3, above=.1]{$rem_1:1$} (s4)

		(s6) edge[bend left=20,sloped,anchor=south, red] node [pos=0.5, above=.1]{$rem_1:1$} (s5)

		(s7) edge[bend left=20,sloped,anchor=south, red] node [pos=0.5, above=.1]{$rem_1:1$} (s6)
	;

	
%
%
	
	\begin{customlegend}[
		legend entries={ 
			$add$,
			$remove$,
			$no\_op$
		},
	legend style={at={(5.5,-2))},font=\footnotesize}] 
	    \addlegendimage{-,blue,thick}
	    \addlegendimage{-,red,thick}
	    \addlegendimage{-,black,thick}
	\end{customlegend}
\end{tikzpicture}
\caption{Example of examined transitions in a MDP solver.} \label{fig:model_steps}
\end{figure}

The direct methods analyze MDPs per se. In our approach we use the PRISM tool to this end. The main challenge here is to define end component states in order to allow for model checking and quantitative verification. Put simply, an end component is one or more connected states, such that no other external state is reachable from them. In our model, every state can be considered as an end component if it is reached through a $no\_op$ action; in other words, when a $no\_op$ action takes place, the model checking terminates. In addition, once the first action is an $add$ (resp. $rem$) one, we allow only VM additions (resp. removals), i.e., we do not allow arbitrary state transitions, which are meaningless in practice. Consequently, every accessible state is visited  and its reward is computed only once. Essentially, the MDP solution examines arbitrary sequences of either additions or removals.
Figure \ref{fig:model_steps} depicts a two-step transition $s_4 \xrightarrow{add_2} s_6 \xrightarrow{add_1} s_7$, using a model similar to the one in Figure \ref{fig:multi_cluster} and assuming that no more than two VMs can be added in every step. Through the investigation of multiple transitions, we are able to compute state's $s_7$ reward ($r_{s_7}$), which may be higher than $r_{s_4}$ and $r_{s_6}$ so take the decision to add 2 new  VMs, even if $R_{s_6} < R_{s_4}$ to claim any potential profits in the near future (the same also applies for down-scaling).

\subsubsection{State Reward Specification}

In our approach, all state rewards are derived from clustering log measurements of similar past conditions, where the similarity is according to the external load. More specifically, we take the approach in \cite{tsoumakos2013automated} as the baseline one: we group log measurements for a specific number of active VMs by their incoming load $\lambda$, and then those measurements are fed into a $k$-means clusterer, which returns $k$ center points in a $n-$dimensional space. $n$ is the number of log measurements that are used to compute the reward; e.g., if the state reward depends on the latency and throughput, then $n=2$.
The center of the biggest region cluster is the one which is selected as the most representative point for every state. However, when reaching such a state as a result of an elasticity action, the real state encountered may be closer to one of the remaining center points. To overcome this concern, we can extend the model so that it explicitly covers all the returned $k$ centers with probabilities that are proportional to the size of their clusters; this is exactly what the model in Figure \ref{fig:multi_cluster} does. Based on the extended model, we can define one state for each of the $k$ clusters.

For the model in Figure \ref{fig:model_original}, where each state may correspond to widely different behaviours,  we distinguish between the following two main approaches: \textit{MB} (mode behaviour), where the state's reward is computed according to the center of the biggest cluster of log measurements, and \textit{EB} (expected behaviour), where we consider all the clusters of log measurements and the reward of each cluster is weighted by its size in order to compute the final aggregate state rewards. The default value of $k$ is 4.

\subsubsection{Other aspects}
Orthogonally to the MDP analysis, we have two options regarding the way we instantiate the model. We can either use the value of the monitored incoming load in each step, or we can use the average value of the incoming load given a sliding window with recent measurements. The rationale behind the usage of a smoothing window is to tackle sudden and temporal peaks of the system's load, which trigger suboptimal elasticity actions.
Those actions are suboptimal because by the time the system settles down and stabilizes again after the elastic action, the peak load no longer exists, so the change in the number of active VMs yields no real benefit. To avoid such early change state decisions, we can use a smoothing window.

Another aspect of the decision making policy is to prohibit the system to continuously take actions with small expected benefits with a view to reduce the probability the system to behave in an unstable manner. The benefit of an elasticity action in each step is defined as the relative difference between the actual value of the utility function given the current system measurements and the expected value if the action takes place (provided through the online model verification). In a post-processing step,  we can enforce only the elasticity actions whose benefit exceeds a user-defined threshold. 

\subsection{Utility Functions}
\label{sec:model_utility}
The actual decision making is based on a given utility function, the maximization of which constitutes the main objective of the decision making module. The rationale of the model is 1) to consider such functions, as functions of the number of active VMs, and 2) to use the utility functions to derive the state rewards in each model instantiation. $Thr$ (for throughput) and $lat$ (for latency) variables are the most significant ones in order to quantify performance and monetary cost, in the way those are employed in the utility functions. However, the model is extensible and it can associate additional variables (e.g., CPU utilization) and utility functions. The current work examines the following utility functions, but the methodology presented above is independent of specific utility functions:
\begin{itemize}
\item $r_1(s) = \begin{cases}
thr/vms\_num &\mbox{if } lat \le x \\
-1 & \mbox{if } lat > x.
\end{cases}
$
\item $r_2(s) = \begin{cases}
1/vms\_num &\mbox{if } lat \le x \\
-1 & \mbox{if } lat > x.
\end{cases}
$
\end{itemize}

In the above functions, $x$ is a latency threshold, which should not be exceeded.
The first utility function $r_1(s)$ tries to maximize the system's throughput, keeping the number of the active VMs as low as possible, taking into consideration a given latency threshold $x$. If the latency threshold is violated the utility function is punished with a negative value -1. The second utility function $r_2(s)$ ignores the system's throughput and only tries to satisfy the latency constraint keeping the number of active VMs to the lowest possible number. $r_2$ utility function is sensitive to the number of latency violations as it is punished with -1 on every violation, while its value is bounded within the range $[1/max\_num\_vms,1/min\_num\_vms]$. It is expected to fit better in a more unstable environment, where the main objective is restricted to constraint satisfaction. In both the utility functions, the number of the VMs is placed to the denominator so that higher utility values correspond to lower over-provisioning, especially for the $r_2$. Any type of costs, overheads or threshold violations, as in the examples, can be modelled as negative rewards in a straightforward manner. Action rewards, despite the fact that are supported by our MDP models, are not considered and will be examined in future work.

\subsection{Quantitative Analysis}
\label{informed_decision}
Probabilistic Computation Tree Logic (PCTL), encapsulated in the PRISM tool, allows for probabilistic quantification of described properties. The primary usage of PCTL in our approach is to extract the maximum expected reward for every state in order to drive elasticity actions. Then, we choose the most profitable state, i.e., the whole problem is a max-max one. However, using PCTL formulae, the users can input additional high-level queries about the probability of the amount of additional resource metrics taking into consideration applied actions and reached states. For example, we can pose questions like  the following:
\emph{``What is the maximum probability of the latency to be less than 30 milliseconds after state $s_7$ is reached?"}, which, in PRISM, can be formulated in this way:
\(Pmax=?\ [ F\ latency<30\ \&\ vms\_num=7 ] \),  where $F$ implies the satisfaction of the reachability property \cite{FKNP11}. Another example question is: \emph{What is the probability that the system will remain in the state decided (assuming that the current environmental conditions do not change)?}

Similarly, we can ask about minimum probabilities and any other metrics used in the model (e.g., throughput). In summary, we can pose any query involving maximum and minimum probabilities and/or rewards. In this way, the user can be more informed about the reason the selected decision was taken and has the ability to examine any metrics of the system, provided that they are employed in the utility functions and thus are captured by the model.

\section{Incorporation Into Existing Systems}
\label{sec:systems}

\begin{figure}[tb!]
\begin{center}
\includegraphics[width=1\columnwidth]{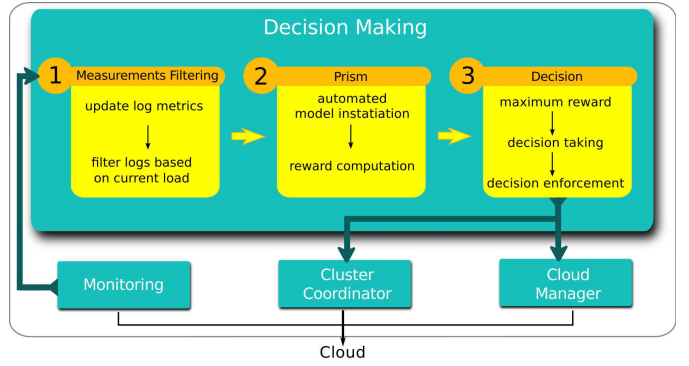}
\end{center}
\caption{Integrating Prism-based decision making with TIRAMOLA.}
\label{fig:tiramola_prism}
\end{figure}

Our elasticity decision approach can be encapsulated in every elastic manager provided that the latter meets the following requirements: it is capable (i) of collecting log measurements that are used for the training and the instantiation of the models and (ii) of enforcing the elasticity decisions taken.

Figure \ref{fig:tiramola_prism} shows how PRISM-based decision making is incorporated within Tiramola \cite{tsoumakos2013automated}, as in our prototype implementation.  Tiramola is a modular, cloud-enabled, open-source system that enables elastic scaling of NoSQL clusters according to user-defined policies and incoming load. It allows seamless interaction with multiple IaaS platforms, requesting/releasing VM resources and orchestrating them inside a NoSQL cluster. Our approach is also compatible with cloud managers like the ones used by Amazon. In that case, the log measurements are provided through Amazon's EC2 {\it CloudWatch} and the decisions can be enforced through Amazon's EC2 {\it Auto Scaling} service. Note that the main current elasticity policy of Amazon is rule-based; in the next section, we compare the efficiency of rule-based decision policies against ours.

\section{Evaluation of Decision Policies}
\label{sec:exps}

The main purpose of this section is to assess the efficiency of the decision policies enabled by our approaches. Since there can be too many combinations of models and decision policy configurations, we compare only a representative subset of decision policies:
\begin{itemize}

\item {\it RE}, which aims to reproduce pure reactive rule-based decision policies, where elasticity actions are triggered by constraint violations, like those enabled by Amazon.

\item {\it RL-MB,} which employs the model in Figure \ref{fig:model_original}, the Q-learning reinforcement learning approach and the \textit{MB} reward specification option (thus reproducing the approach in \cite{tsoumakos2013automated}, which represents the state-of-the-art in NoSQL elasticity).

\item {\it MDP-MB,} which differs from \textit{RL-MB} in that a direct MDP solver is used through PRISM.

\item {\it MDP-EB,} which differs from \textit{MDP-MB} in that it employs the \textit{EB} reward specification option.

\item {\it MDP2,} which employs the model in Figure \ref{fig:multi_cluster} and a direct MDP solver is used through PRISM.

\item {\it MDP3,} which combines \textit{MDP2} with the model in Figure \ref{fig:looking_forward}.

\end{itemize}

\begin{figure}[tbH]
\centering
\subfloat[Real Dataset]{\includegraphics[width=2.8in]{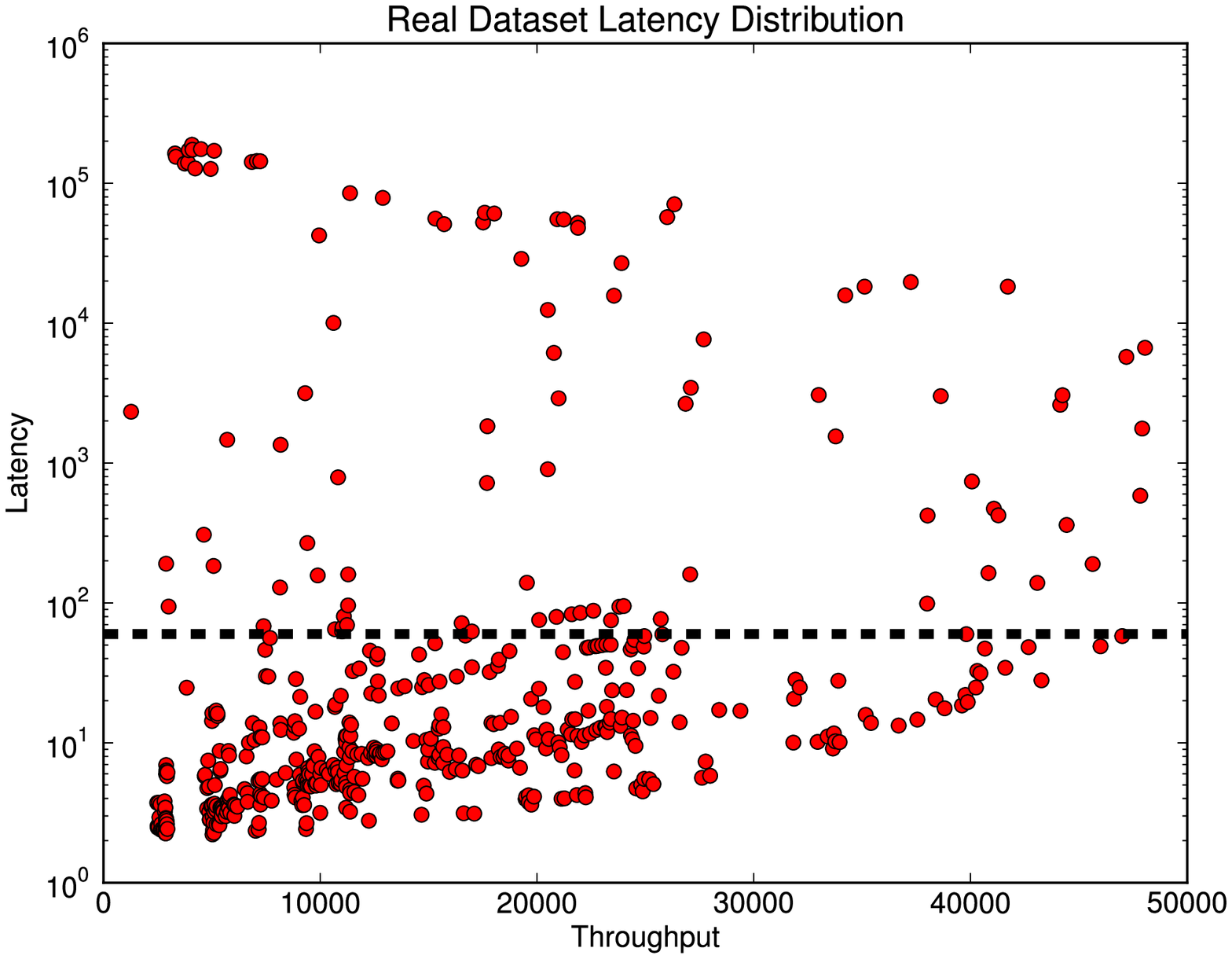}\label{fig:lat_dist_d1f}}

\subfloat[Synthetic Dataset]{\includegraphics[width=2.8in]{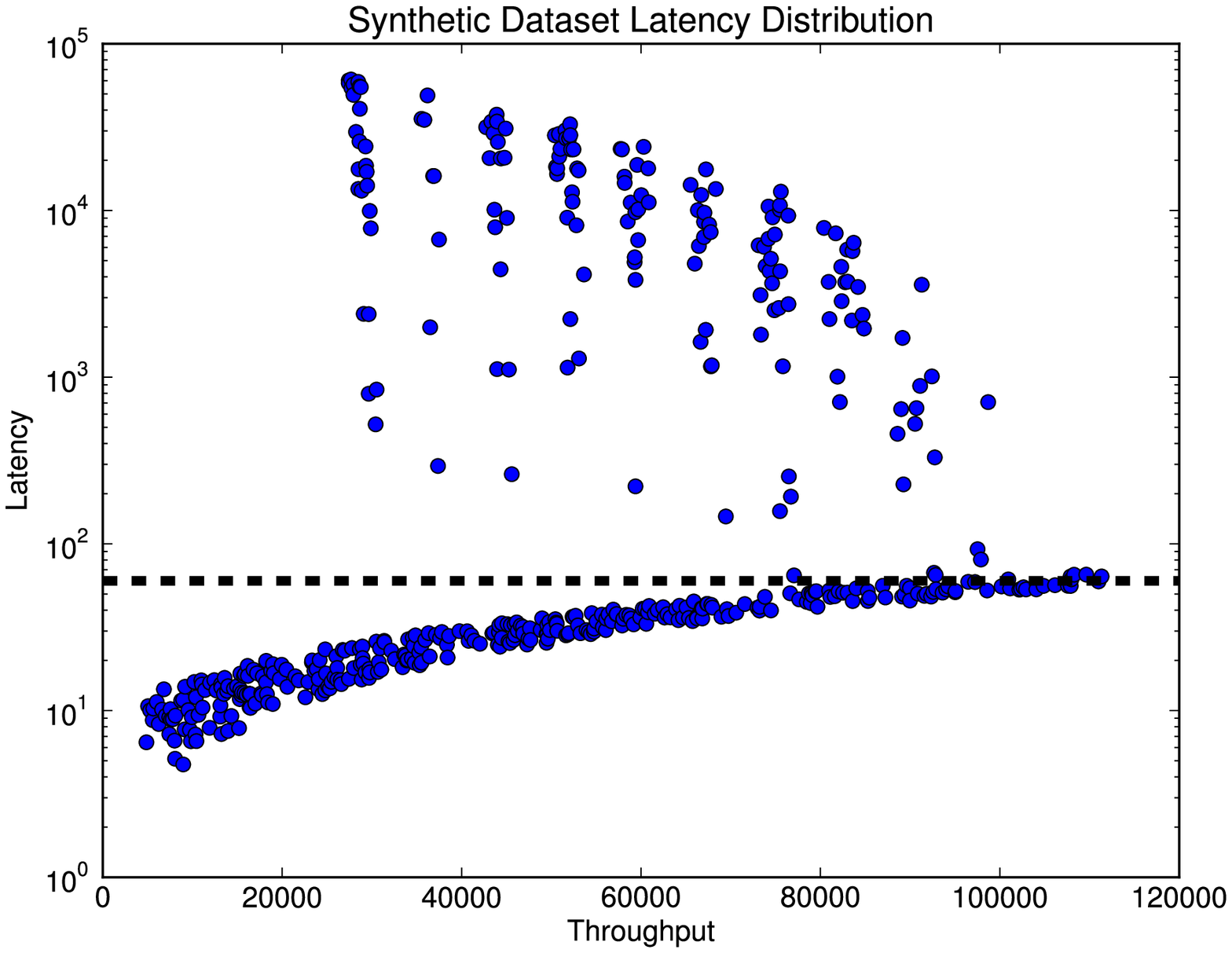}\label{fig:lat_dist_gf}}
\caption{Datasets Latency Distribution}
\label{fig:lat_dist}

\end{figure}

We ran two main sets of experiments. In both sets the incoming load of the system was being modified in a sinusoidal fashion, as in a standard seasonal workload pattern according to \cite{qureshi2009cutting,chen2008energy}. In the first set, we use real data about incoming load and system latency and throughput from an elastic NoSQL cluster. This set is characterized by unpredictable and widely variant behaviour for the same values of incoming load. Figure \ref{fig:lat_dist} illustrates the latency distribution for VM sizes and values of load. The second dataset is synthetically generated and dynamically evolving as well, but the system behaviour is less unpredictable. For both datasets, we have used two load variations (denoted as LV1, LV2), where the external load begins from the minimum and the average value (through a $\pi/2$ shift), respectively.

To assess the efficiency for each of the 5 policies above, we measure the actual value of the utility function in each time unit; time units will be explained later. We use the two utility functions presented in Section \ref{sec:model_utility}. Note that during the MDP analysis, these functions are used to derive the state rewards; however, the expected rewards may diverge from the actual system properties (e.g., latency and throughput). Also, the ratio of different utility values does not imply relative performance, and thus we do not use it as a comparison metric. However, the utility values combined with the number of latency violations, can indicate system over-provisioning. By default, we allow up to 3 VMs to be added and 2 to be removed in a single action, and we do not employ smoothing windows and/or benefit thresholds. The (upper) latency threshold is set to 60msecs, and is depicted as a dotted line in Figure \ref{fig:lat_dist}; as shown in the figure, a big portion of system configurations do not meet the constraint.
The cluster is initialized with 4 VMs which is the minimum number of VMs, as it is set in our experiments. The implication is that for LV1, no urgent elasticity actions are required in the beginning, contrary to the situation in LV2.
In addition, we measure the time overhead in reaching decisions.
To get as close to Amazon's decision policy as possible, {\it RE} makes actions of only a pre-specified size. I.e if the add limit is 4, then, whenever an increase is decided, 4 VMs are going to be added. In addition, {\it RE} requires a second, lower latency threshold, which is set to half the upper bound, i.e., 30msecs.

\subsection{Experiments with Real Data}

\subsubsection{Experimental Setup}
In order to collect real data, we conducted log measurement experiments using the OKEANOS IaaS infrastructure \cite{KoukisVK13}, and the YCSB benchmark.
For our NoSQL cluster, we have used 4 client VMs as load generators with 2 VCPUs and 2GB of RAM each, and up to 16 servers VMs with 2 VCPUs, 4GB of RAM and 20GB storage each. The volume storage service supporting each server VM utilizes RADOS, the distributed object store underlying the Ceph parallel filesystem.
Hbase NoSQL DB version 0.94.11 is installed and configured on every server VM atop of Hadoop version 1.0.4. A heavily modified version of YCSB-0.1.3  ran on every client VM to produce the load; the modifications were made to support database metrics reporting on ganglia \cite{massie2004ganglia}. Using YCSB, we have created 10 million rows (approx. 10GB) of data to the Hbase NoSQL DB with replication factor 2. The workload consists of asynchronous read requests in uniform distribution. Based on the observations of  \cite{tsoumakos2013automated}, we have enabled region rebalancing without data rebalancing (neither Hbase compaction nor HDFS balancer).
We have created varying load by modifying the target and threads parameters of the YCSB tool, 
producing load from 1000 (req/sec) up to 46000 (req/sec) with a step of 1000 (req/sec). We collected measurements every 30 secs, and in each sine period, there were 315 measurements.

The collected measurements are used firstly, to populate the initial logs of each policy, and secondly, to emulate a real situation. Through emulation, we managed to fairly test each policy on an equal basis, which could not be done if each policy ran separately in a real cluster. In our emulation, a  time unit corresponds to the measurement collection period, i.e., 30 secs. We allow an elasticity action to take place every 10 time units, to emulate a system that may modify the VMs every 5 mins.  As the emulated load is generated based on the logs which also act as training set, we consider that the system is well trained. In the emulation, for each number of active VMs and external load, we defined the emulated system state after choosing the appropriate log measurements and adding some noise. If the pair of [\textit{active VMs number} - \textit{external load}] is not included in the collected measurements, we deduce such values based on their closest neighbours.

\subsubsection{Experimental Results}

\begin{figure}[!tbH]
\centering
\subfloat[Average Utility for $r_1$]{\includegraphics[width=2.9in]{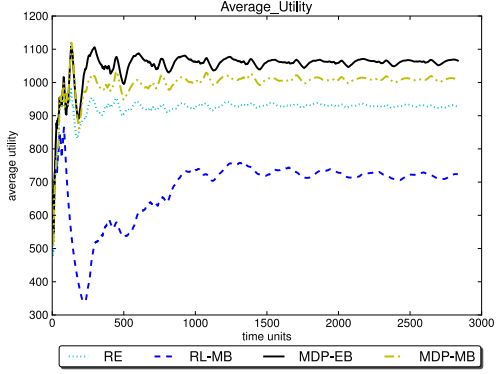}\label{fig:av_rew_d1f_r1_wf}}

\subfloat[Average Utility for $r_2$]{\includegraphics[width=2.9in]{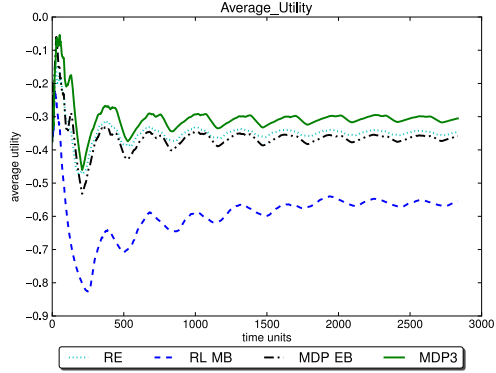}\label{fig:av_rew_d1f_r2_wf}}
\caption{Average Utility for Real Data-LV1}
\label{fig:avg_rew_d1f_d1}

\end{figure}


Figures \ref{fig:av_rew_d1f_r1_wf} and \ref{fig:av_rew_d1f_r2_wf} show the average utility values for LV1, using the $r_1$ and $r_2$ utility functions, respectively. In the figures, for readability reasons, we plot only the {\it RE}, {\it RL-MB}, {\it MDP-EB} and the best performing policy, which is always an MPD-based one and plotted with a solid line . If the best one is {\it MDP-EB}, we choose an additional representative policy.
The figures correspond to averages from 10 runs. MDP-based policies seem to adapt better in both scenarios and yield higher utility values. In general, the differences between the utility of the MDP policies are small.
{\it RE} does not perform that well, as it is only based on latency thresholds. However, {\it RE} adapts better than the {\it RL-MB} policy, which does not succeed in adapting efficiently in such an unstable environment.

\begin{figure}[!tbh]
\centering
\includegraphics[width=2.65in]{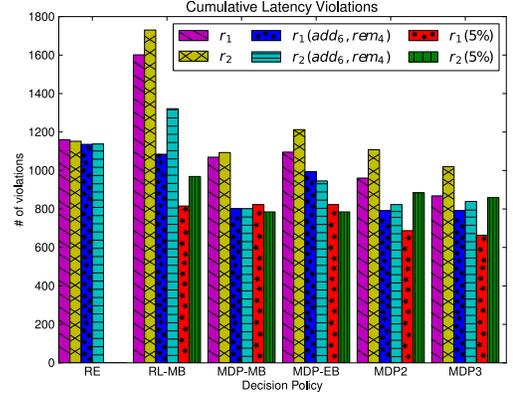}
\caption{Latency Violations for Real Data-LV1}
\label{fig:lat_vio_d1f}

\end{figure}

The same conclusions can be also drawn from Figure \ref{fig:lat_vio_d1f}, where the average cumulative number of latency constraint violations are illustrated for both $r_1$ and $r_2$ utility functions (see the two leftmost bars in each group). MDP-based policies result in $~37\%$ less violations on average than {\it RL-MB}. {\it MDP3} can yield $45\%$ less violations than {\it RL-MB}, for $r_1$.

\begin{figure}[!tbh]
\centering
\subfloat[Average Utility for $r_1$]{\includegraphics[width=2.9in]{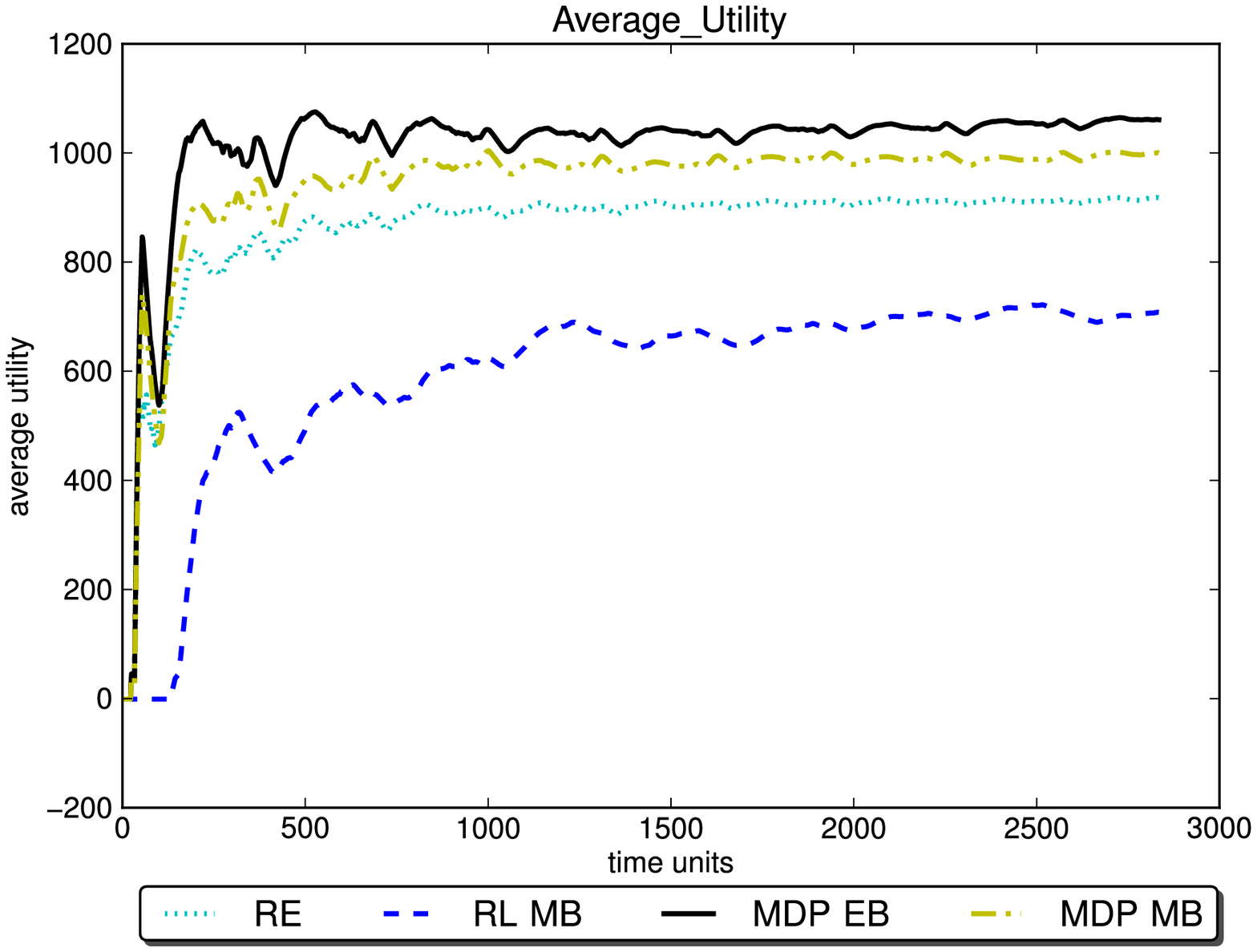}\label{fig:av_rew_d1_r1_wf}}

\subfloat[Average Utility for $r_2$]{\includegraphics[width=2.9in]{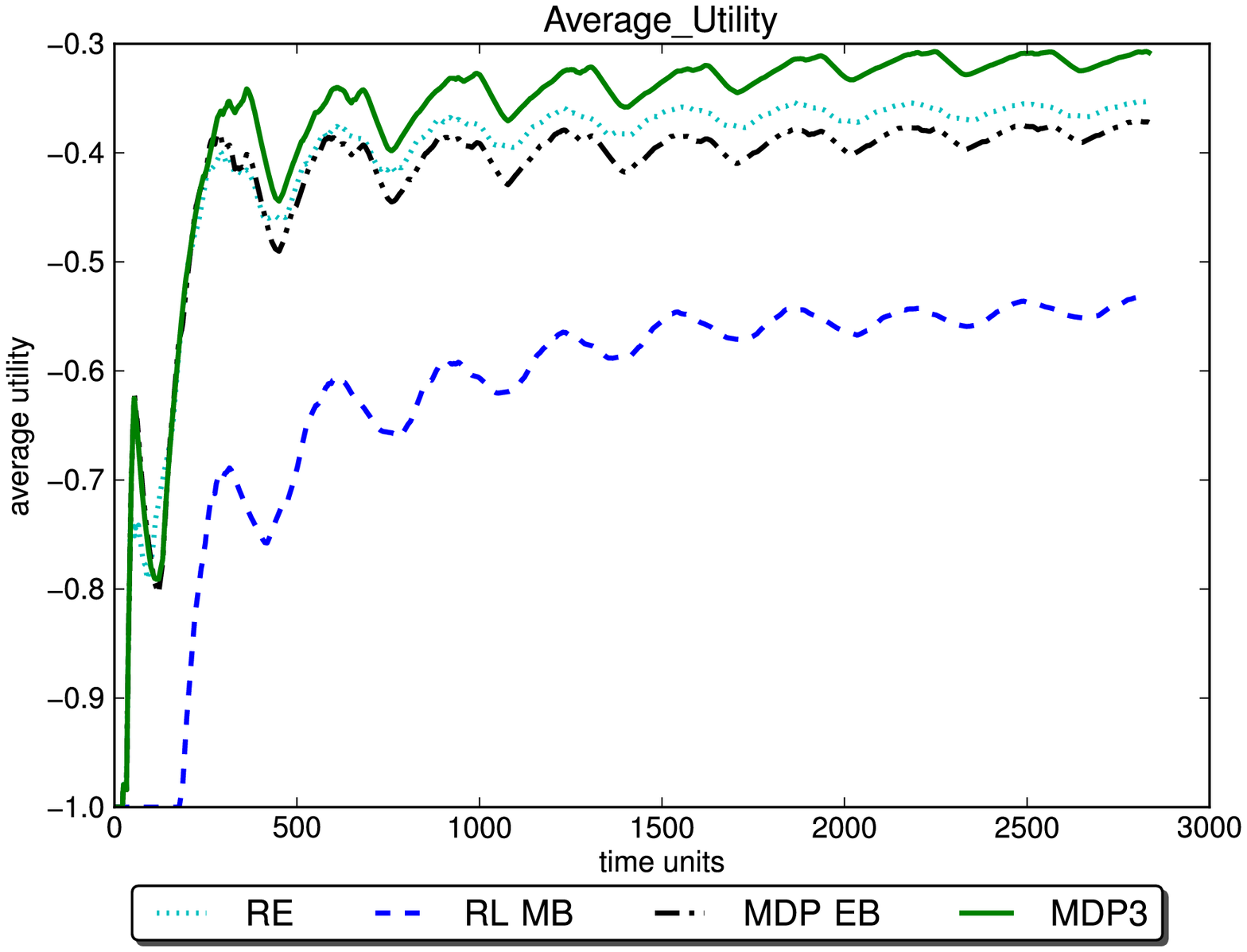}\label{fig:av_rew_d1_r2_wf}}\\
\caption{Average Utility for Real Data-LV2}
\label{fig:avg_rew_d1}

\end{figure}

\begin{figure}[!tbh]
\centering
\subfloat[RE]{\includegraphics[width=2.8in]{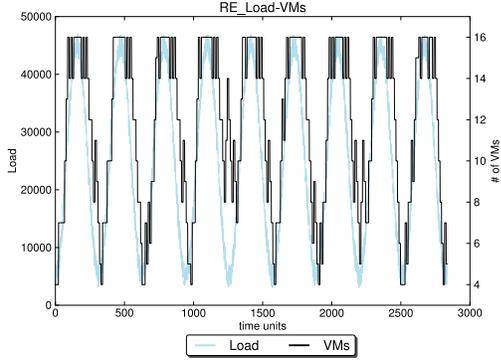}\label{fig:load_vms_re_d1f_r1}}

\subfloat[RL-MB]{\includegraphics[width=2.8in]{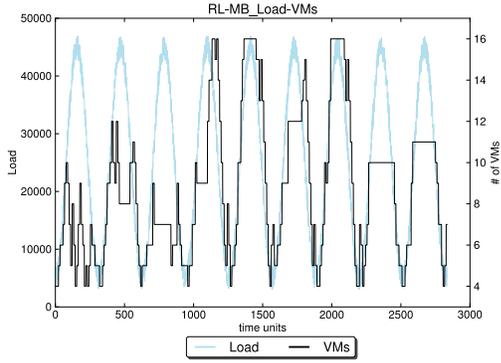}\label{fig:load_vms_rl_d1f_r1}}

\subfloat[MDP-EB]{\includegraphics[width=2.8in]{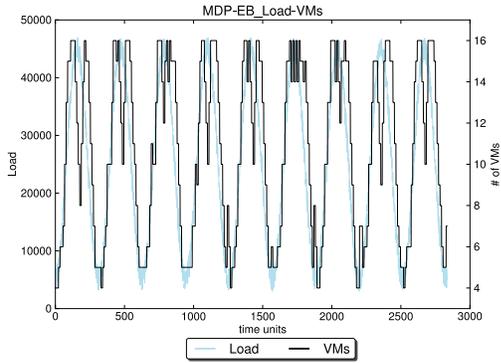}\label{fig:load_vms_mdpeb_d1f_r1}}
\caption{External Load and VMs for Real Data-LV1}
\label{fig:load_d1f_r1}

\end{figure}

\begin{figure}[tbH]
\centering
\subfloat[$r_1$ - Real Data-LV1]{\includegraphics[width=3in]{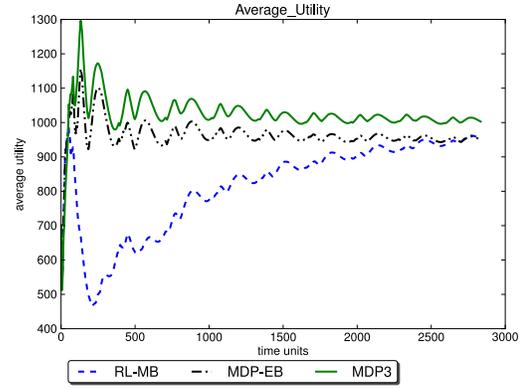}\label{fig:av_rew_d1f_r1_wf_5p}}

\subfloat[$r_2$ - Real Data-LV1]{\includegraphics[width=3in]{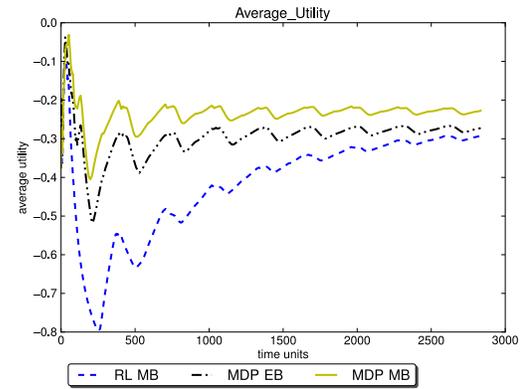}\label{fig:av_rew_d1f_r2_wf_5p}}
\caption{Applying a benefit threshold of $5\%$.}
\label{fig:avg_rew_d1f_6a_4r_5p}

\end{figure}

For LV2, the same observations regarding the superiority of MDP-based policies hold (see Figures \ref{fig:av_rew_d1_r1_wf}-\ref{fig:av_rew_d1_r2_wf}). 
Figure \ref{fig:load_d1f_r1} shows indicative actions for each policy in 1 of the 10 runs, when the utility function is $r_1$ (the actions for $r_2$ are similar). We only provide a trace for {\it MDP-EB} due to space limitations. As we observe, we continuously perform elasticity actions because of the constantly evolving external load.
{\it MDP-EB} better follows the incoming load, whereas other policies, such as {\it RL-MB} cannot do so mostly because of the fact that, especially for large values of load, most of the corresponding log measurements violate the latency threshold, regardless of the VM size (not explicitly shown in Figure \ref{fig:lat_dist}).

In Figure \ref{fig:av_rew_d1f_r1_wf}, we observe that \textit{RE} policy achieves a smaller reward than the \textit{MDP}-based policies, but higher than the \textit{RL-MB} policy. In Figure \ref{fig:lat_vio_d1f} we observe that  \textit{RE} and \textit{MDP-EB}, which is the best in that experiment, have almost equal number of  latency violations. This raises the question as to why there is so much difference in the average utility value. The answer is provided in Figure \ref{fig:load_d1f_r1}. As we can see, \textit{RL-MB} under-provisions (provides less VMs than necessary) the system in most of the cases, whereas \textit{RE} does not avoid over-provisioning. This explains the overall superiority of \textit{MDP-EB}.

\begin{figure*}[!tbH]
\centering
\subfloat[$r_1$ - Real Data-LV1]{\includegraphics[width=2.8in]{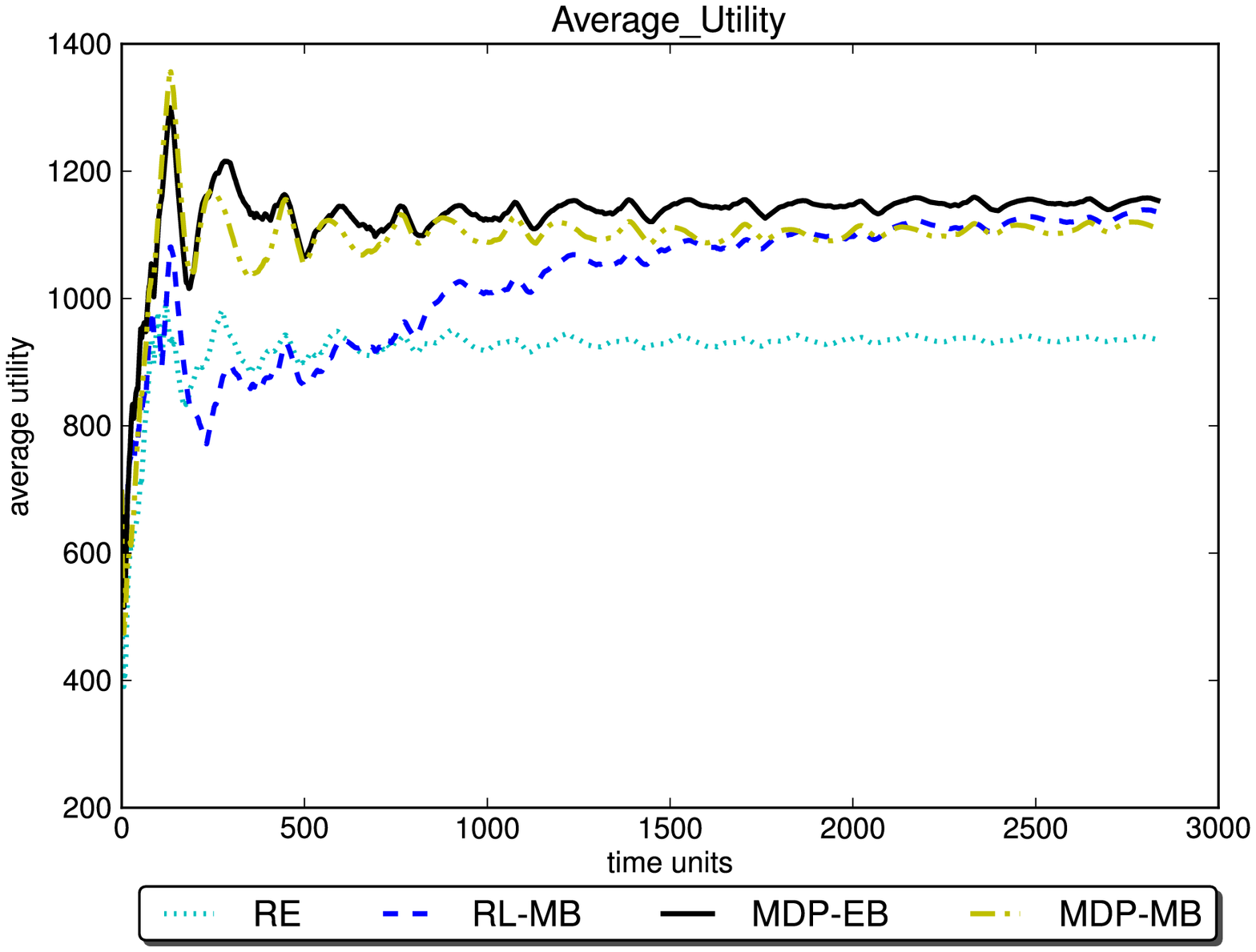}\label{fig:av_rew_d1f_r1_wf_6a_4r}}
\subfloat[$r_2$ - Real Data-LV1]{\includegraphics[width=2.8in]{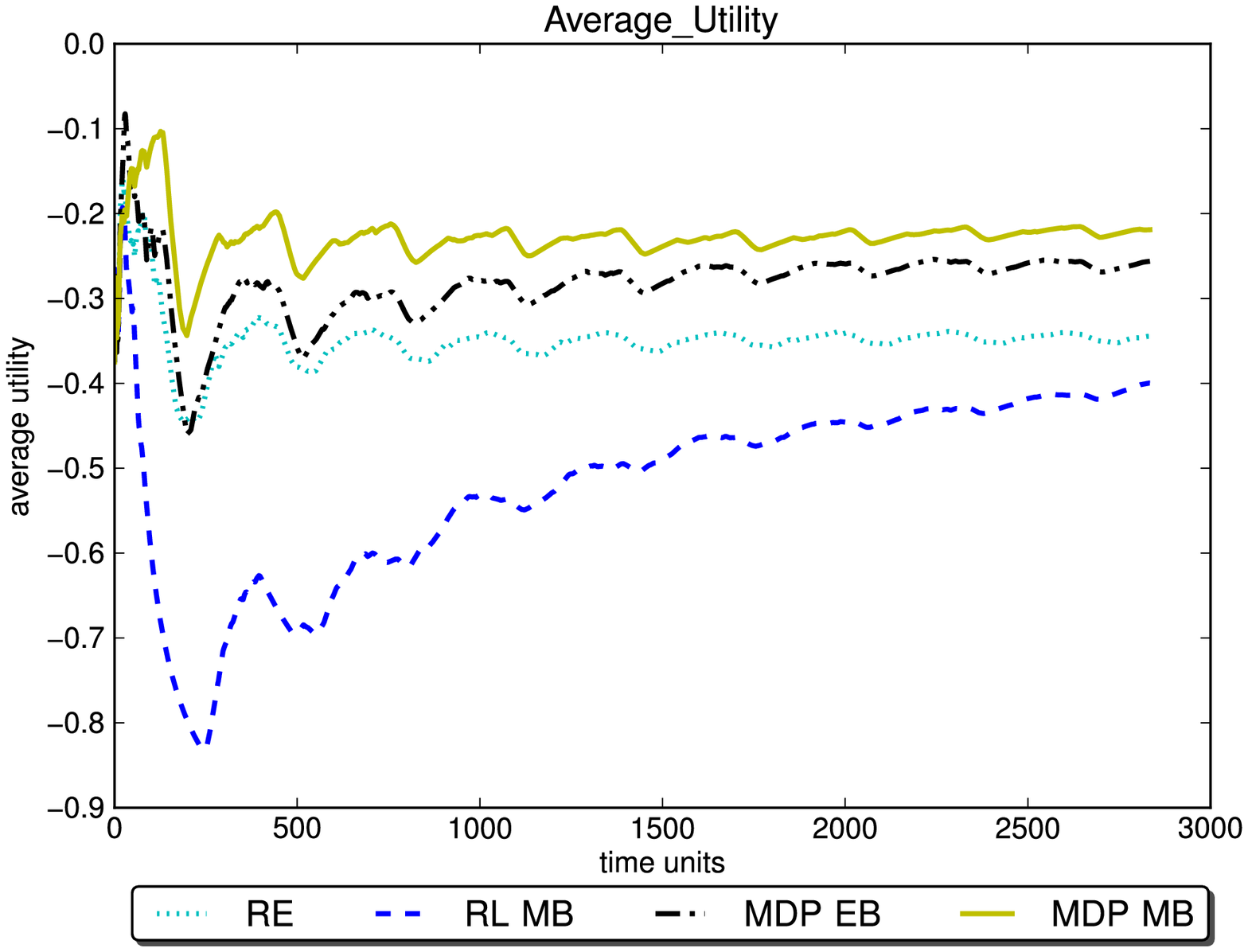}\label{fig:av_rew_d1f_r2_wf_6a_4r}}

\subfloat[$r_1$ - Real Data-LV2]{\includegraphics[width=2.8in]{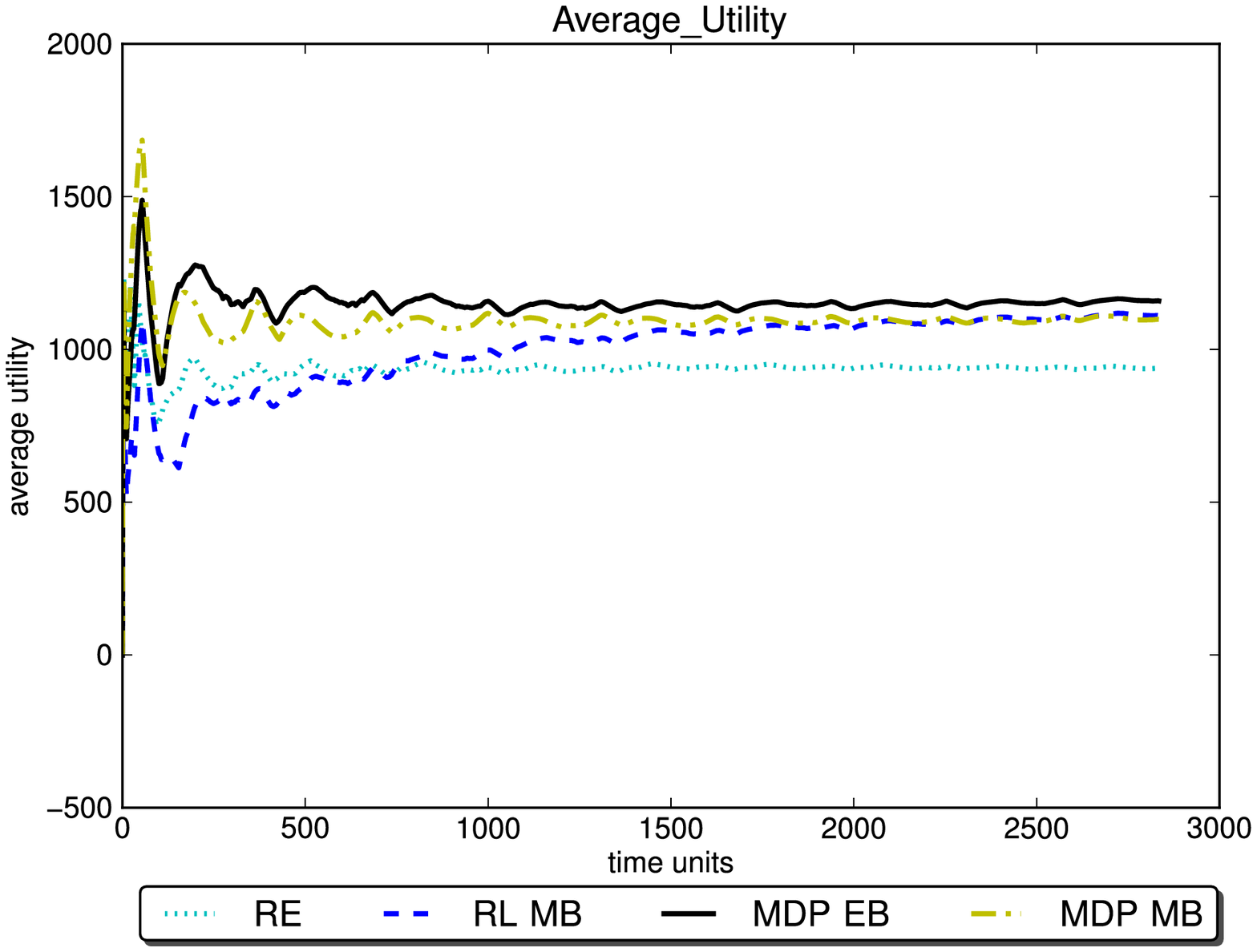}\label{fig:av_rew_d1_r1_wf_6a_4r}}
\subfloat[$r_2$ - Real Data-LV2]{\includegraphics[width=2.8in]{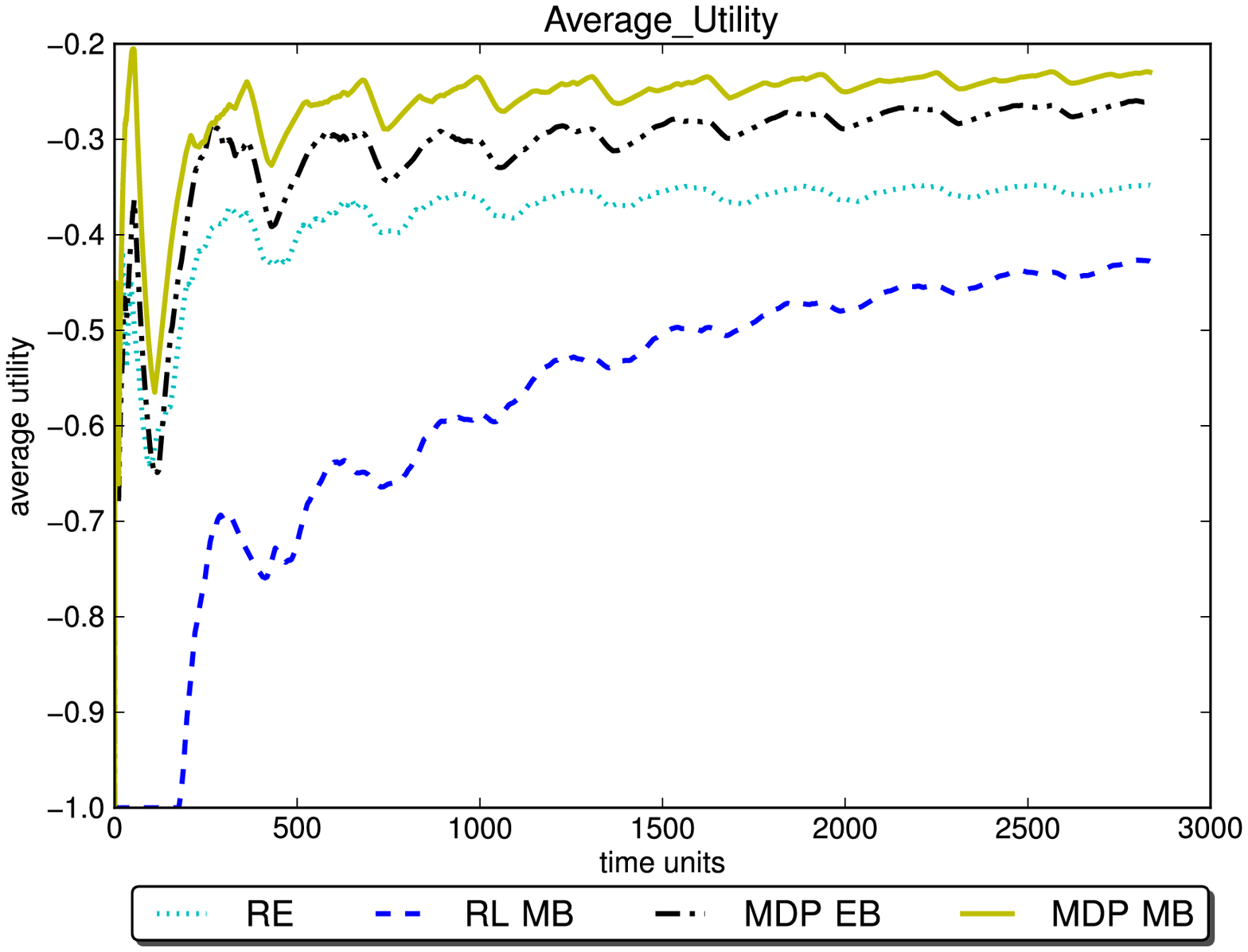}\label{fig:av_rew_d1_r2_wf_6a_4r}}
\caption{Allowing up to 6 (resp. 4) VMs to be added (resp. removed) }
\label{fig:avg_rew_d1f_6a_4r_5p}
\end{figure*}

We also consider cases with increased add and remove limits, softening the system constraints. In this experiment, we allow up to 6 additions (about $1/3rd$ of the total number of VMs) and up to 4 removals ($1/4th$ of the total VMs) in a single step. Figures \ref{fig:av_rew_d1f_r1_wf_6a_4r} and \ref{fig:av_rew_d1f_r2_wf_6a_4r} (to be compared with the Figures \ref{fig:av_rew_d1f_r1_wf} and \ref{fig:av_rew_d1f_r2_wf}) show that all the policies, and especially {\it RL-MB}, are improved due to this configuration. The cumulative latency violations are also decreased, e.g., up to $32\%$ for {\it RL-MB}, as shown in the two middle bars in  Figure \ref{fig:lat_vio_d1f}. For LV2, the increase of the number of allowed added or removed VMs allows {\it RL-MB} to catch up with the sudden need for resources, as shown in Figure \ref{fig:av_rew_d1_r1_wf_6a_4r}. Still, MPD-based policies are superior, especially for $r_2$.

The addition of a benefit threshold (set to $5\%$) can further improve the performance of {\it RL-MB}, as shown in Figures \ref{fig:av_rew_d1f_r1_wf_5p},\ref{fig:av_rew_d1f_r2_wf_5p} (increasing over-provisioning though, which is not shown in the figures). For the  MDP-based policies, this parametrization does not lead to improvements for the $r_1$ utility function. On the other hand, using the $r_2$ utility function combined with the benefit threshold, helps MDP-based policies to further decrease the total number of threshold violations. {\it RE} is not considered in this scope, as it applies its decision without taking into consideration any benefit threshold. The number of the threshold violations are shown in the two rightmost bars in Figure \ref{fig:lat_vio_d1f}. We also experimented with applying a smoothing window, but we did not observe significant differences in the results presented; this can be explained by the sinusoidal incoming load, for which, the most recent value is, in general, more representative than the smoothed average.

\subsection{Experiments with Synthetic Data}

\begin{figure*}[!tb]
\centering
\subfloat[$r_1$]{\includegraphics[width=2.8in]{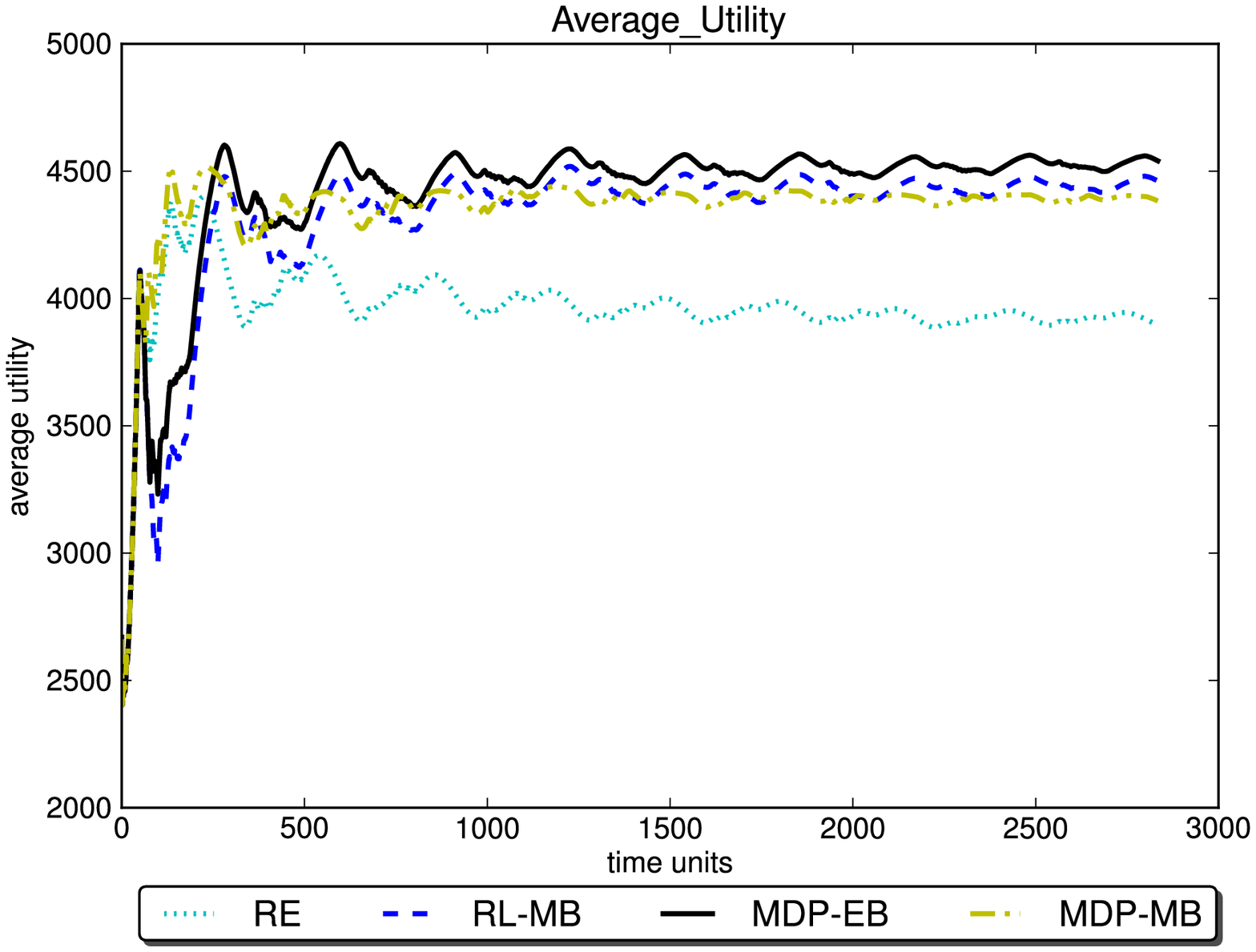}\label{fig:av_rew_gf_r1_wf}}
\subfloat[$r_2$]{\includegraphics[width=2.8in]{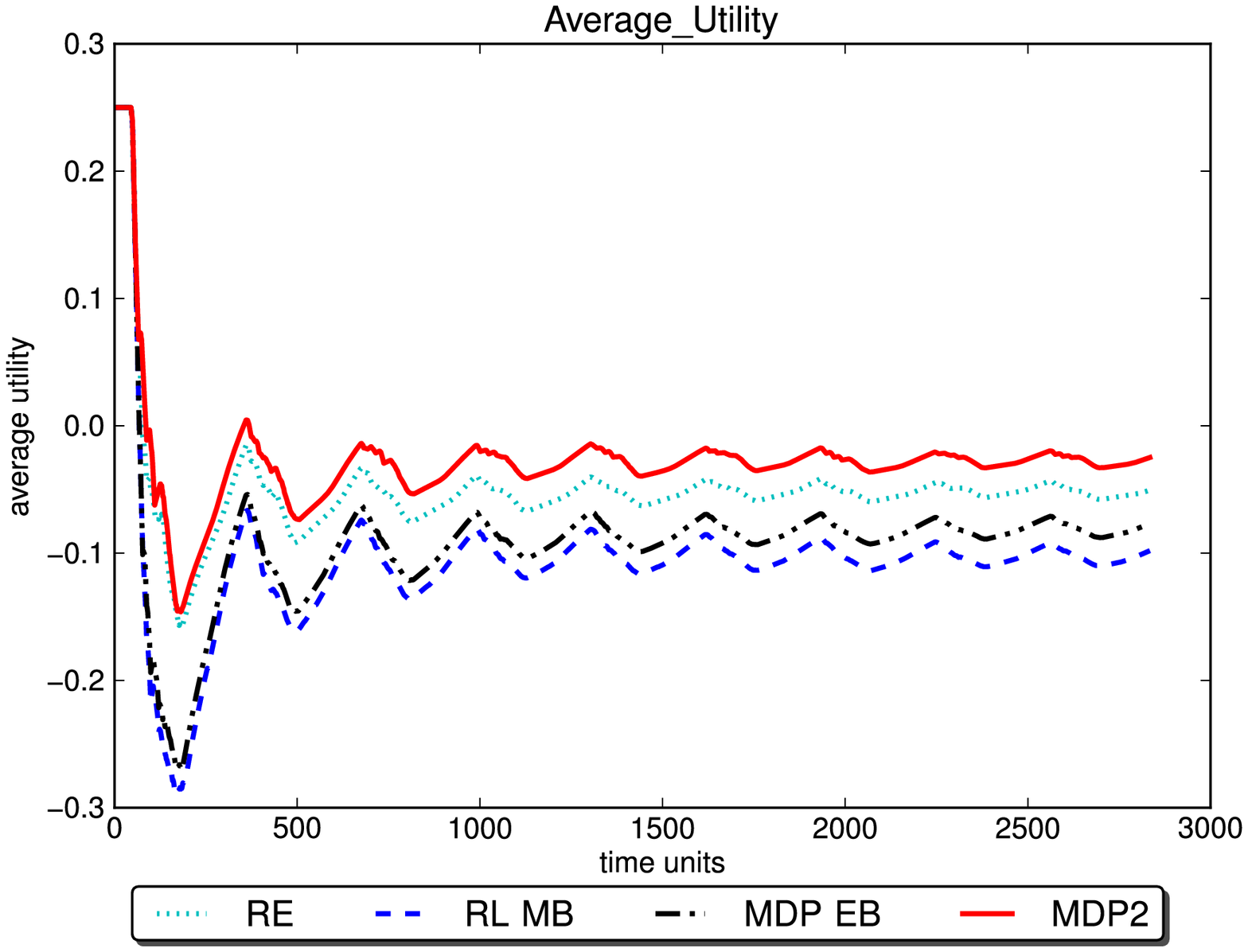}\label{fig:av_rew_gf_r2_wf}}

\subfloat[$r_1$ - 6 additions, 4 removals]{\includegraphics[width=2.8in]{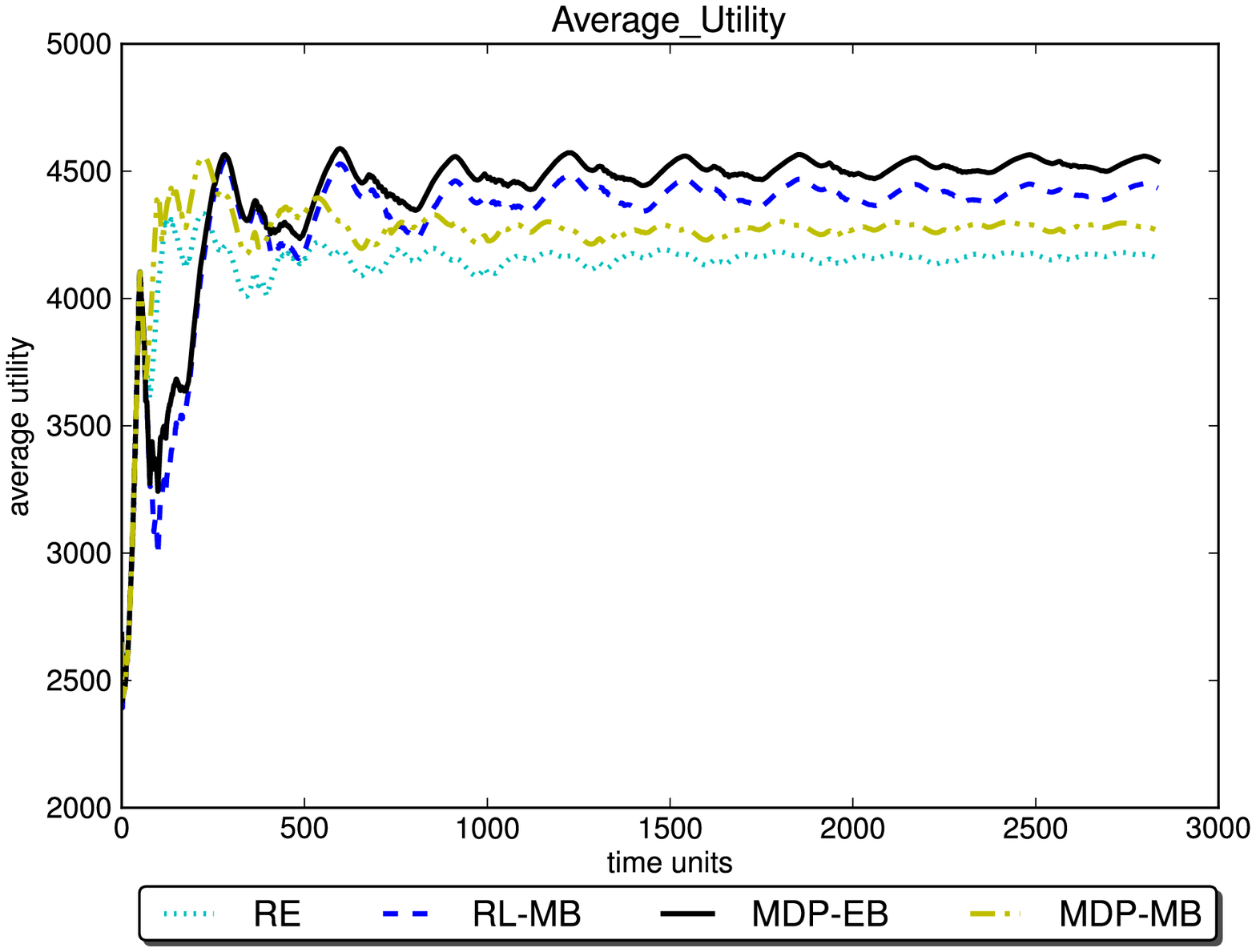}\label{fig:av_rew_gf_r1_wf_6a_4r}}
\subfloat[$r_2$ - 6 additions, 4 removals]{\includegraphics[width=2.8in]{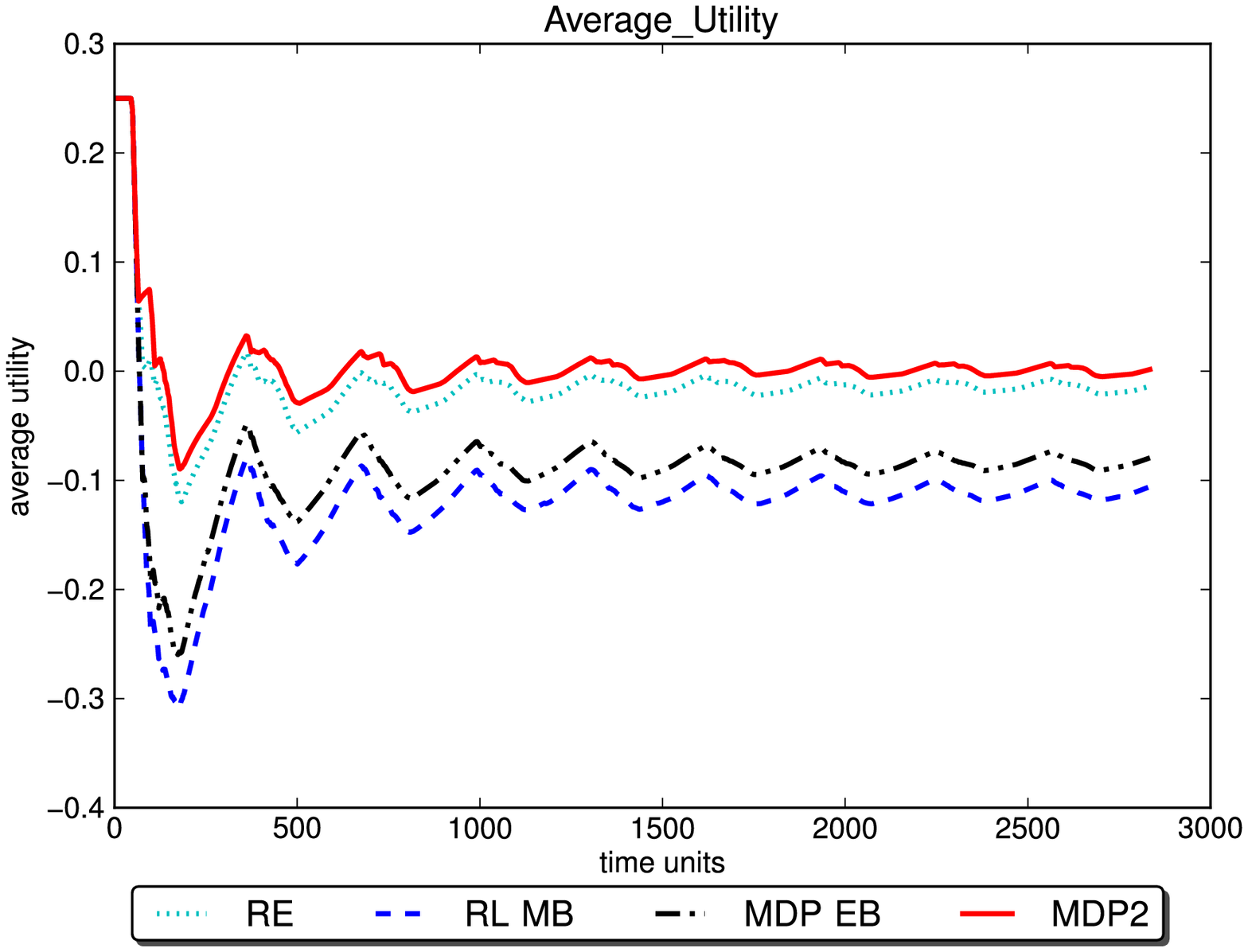}\label{fig:av_rew_gf_r2_wf_6a_4r}}
\caption{Average Utility for Synthetic Data-LV1}
\label{fig:avg_rew_gf}

\end{figure*}

\begin{figure}[tbH]
\centering
\subfloat[RE]{\includegraphics[width=2.8in]{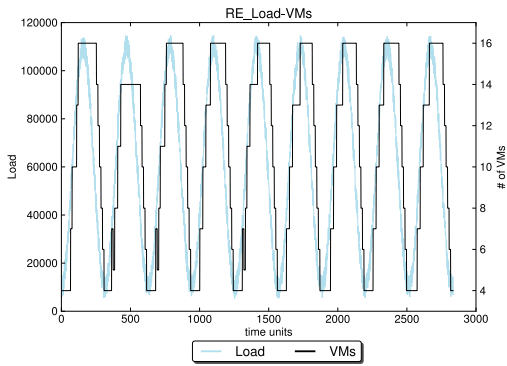}\label{fig:load_vms_re_gf_r1}}

\subfloat[RL-MB]{\includegraphics[width=2.8in]{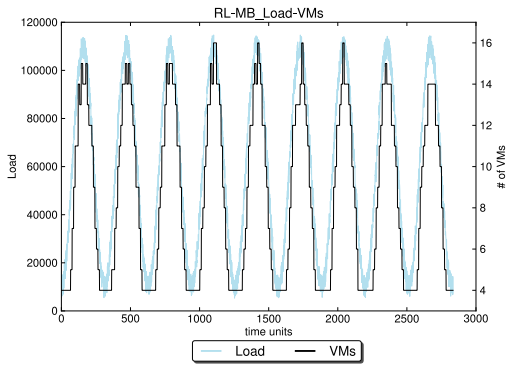}\label{fig:load_vms_rl_gf_r1}}

\subfloat[MDP-EB]{\includegraphics[width=2.8in]{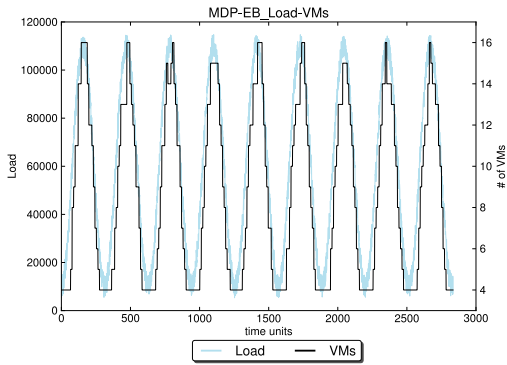}\label{fig:load_vms_mdpeb_gf_r1}}
\caption{External Load and VMs for Synt. Data-LV1}
\label{fig:load_gf_r1}

\end{figure}

\begin{figure}[tb]
\centering
\includegraphics[width=2.65in]{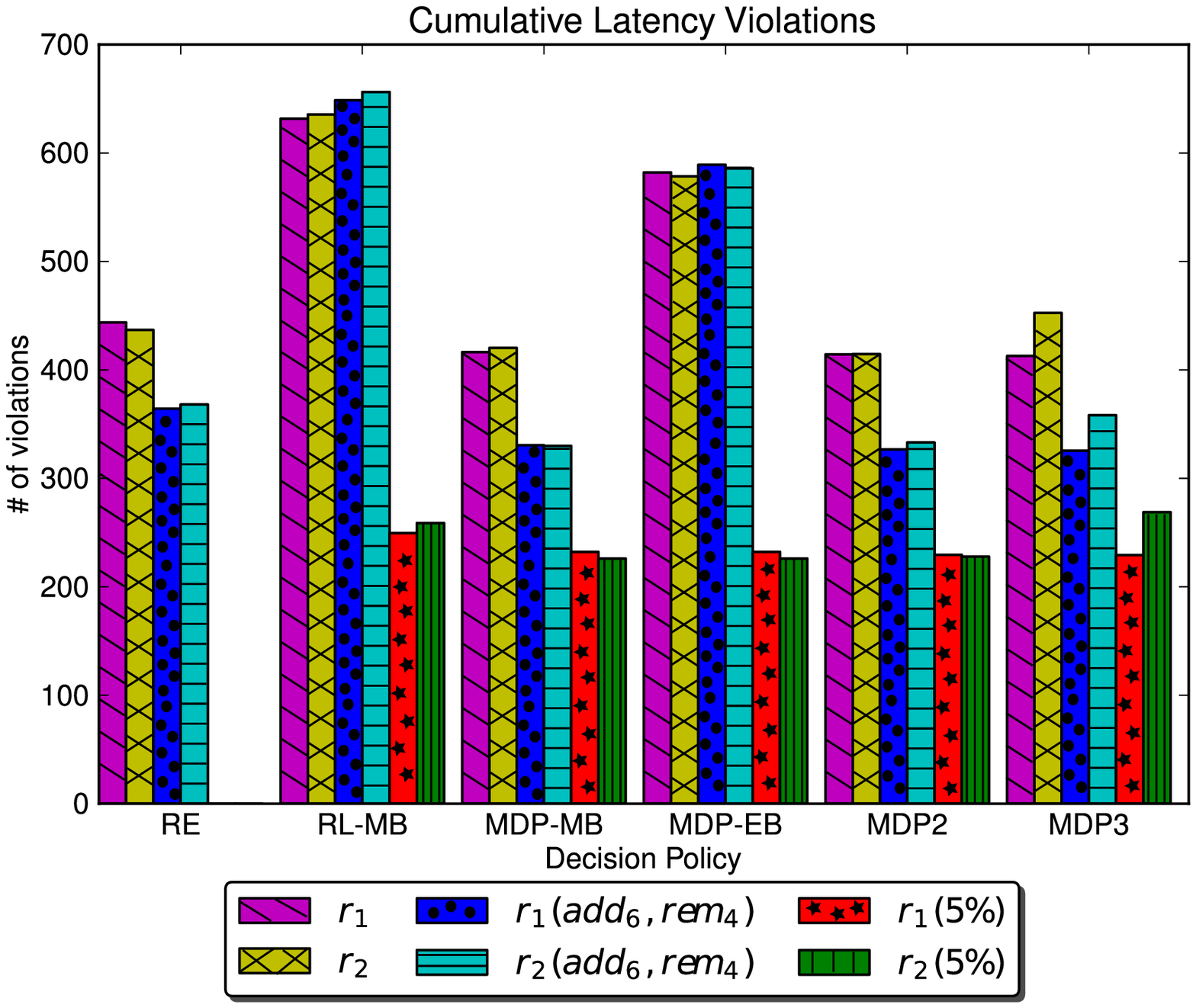}
\caption{Latency Violations for $r_2$ - Synthetic Data}
\label{fig:lat_vio_gf}
\end{figure}

The synthetic load data and the training set are generated in a similar fashion. A load generator produces sinusoidally load and computes throughput and latency values based on the current state.
In the synthetic dataset, the throughput and latency variation is smaller for the same load. Figures \ref{fig:av_rew_gf_r1_wf},\ref{fig:av_rew_gf_r2_wf} show the average utility values for LV1. As we observe, the differences between the decision policies are smaller, especially for the $r_2$ utility function, although the utility values attained by MDP-based policies are still higher than those of other approaches. Concerning the $r_1$ utility function, {\it MDP3} (not plotted) and {\it RE}  act almost the same, as the former takes decisions mostly close to the limits. {\it MDP2} (not plotted) and {\it MDP-MB} also exhibit very similar behaviour, as {\it MDP2} follows paths in the model with higher probabilities and, since this dataset has predictable throughput and latency behaviour, the most probable paths are those from the biggest cluster of log measurements. The same applies to {\it RL-MB} and {\it MDP-EB}, which exhibit similar behaviour. However, since {\it MDP-EB} takes into account more factors for the rewards computation, it adapts better to the load variations. As in the real dataset, the {\it RL-MB} delays adaptation to the load variation for LV2.

Figure \ref{fig:load_gf_r1} illustrates indicative actions for each policy in 1 of the 10 runs for utility function $r_1$ (for $r_2$, actions are  similar). As we see in this figure, the actions are more stable. This is expected as the system behaviour in the synthetic load is more predictable. Figure \ref{fig:lat_vio_gf} shows the number of threshold violations, which can be significantly less for MPD-based policies.
Increasing the allowed additions/removals has a small positive affect on {\it MDP3} and {\it RE} for the $r_1$ utility function (shown in Figure \ref{fig:av_rew_gf_r1_wf_6a_4r}). For $r_2$, the positive impact is slightly more significant, as shown in Figure \ref{fig:av_rew_gf_r2_wf_6a_4r}.

We also examined the impact of the benefit threshold and the smoothing window. On average, applying a benefit threshold leads to an increase in the yielded utilities for all policies, whereas the smoothing window seems not to have any notable impact (no figures are presented). We expect that the smoothing window will perform better in workloads with peaks but this will be further examined in  future work. The general observation that MDP-policies are capable of achieving the highest utility holds.

\subsection{Decision Making Overhead and Discussion}
\label{sec:exps_overhead}
For the experiments, we used a machine with Xeon 3075 CPU and 4GM RAM. On average, the {\it RL-MB} decision policy takes 0.07secs to reach a decision, while MDP-based policies that invoke PRISM take 1.5secs approximately, without significant differences between them (even for the  \textit{MDP3} policy, which employs the largest model). The {\it RE} policy decides almost instantly. The difference of two orders of magnitude  in the running time between the {\it RL-MB} and MDP based decision policies is not significant in practice, since we typically take elasticity actions every 5 or 10 minutes.

Overall, the evaluation results provide strong insights into the capability of MPD-based policies of maximizing user-defined utility functions compared to RL-based and reactive alternatives. There is no clear winner between the 4 MDP-based flavours examined. For $r_1$, which considers both performance-related metrics (such as throughput) and user-defined thresholds, {\it MDP-EB} yields the highest utility values in almost all the cases. However, when the objective is mainly to satisfy a constraint (exemplified by $r_2$), the remaining three flavours perform better than {\it MDP-EB} and, in general, similarly to each other in terms of the utilities achieved. This is a strong indication that the underlying MDP model need not be more complex than the one in Figure \ref{fig:model_original}. However, if only the number of threshold violations per se is considered regardless of the encapsulating utility function, then the more complex model of Figure \ref{fig:multi_cluster} prevails, especially for the real dataset.

\section{Related Work}
\label{sec:rw}

In this section, we briefly discuss works similar to ours, focusing on the elasticity decision making process. A key distinctive feature of our work is that it employs online quantitative verification to drive elasticity, which is novel. In addition, we can classify elasticity decision making as either reactive (e.g., being  triggered when a threshold is exceeded) or pro-active. Our approach falls into the latter category, since it assumes that the system periodically seeks to move to a new state with higher utility, even when there is no constraint violation in its current state.

A work very close to ours is the Tiramola system \cite{tsoumakos2013automated}, which, for deciding the number of VMS, employs periodically activated reinforcement learning on top of MDP models. In this work, we extend this approach as discussed in the previous sections, and we allow for additional MDP solvers, dynamically instantiated models and quantitative analysis. Regarding other forms of elasticity, the work of \cite{shen_cloudscale:_2011} presents an elastic system that can manage VM resource usage according to applied workloads and agreed SLA, where we target the amount of VMs.
PRESS \cite{gong_press:_2010} is another framework which automatically scales VMs according to observed workloads while considering energy consumption and SLA. The framework uses prediction for reducing under and over provisioning errors and use CPU frequency scaling for achieving energy savings with minimal impact on SLA.
In the work of \cite{marshall_elastic_2010}, a model implementing an elastic site resource manager that dynamically consolidates remote cloud resources based on predefined policies is presented.
Nevertheless, their approach targets mostly cluster-based systems.
In \cite{elmroth_self-management_2011} the authors describe a hierarchical control mechanism, with separate controllers for each application tier, each controller having a set of sub low level controllers for memory, storage, bandwidth and CPU.
Wang et al \cite{wang_flexible_2011} focus on software resource allocation (e.g. thread pool size management, DB connection pool, etc.).

The work of \cite{trushkowsky_scads_2011} targets dynamic resource allocation for distributed storage systems, which need to maintain strict performance SLA.
In our case, we are targeting a broader area of applications, resources and available cloud providers. The automatic scaling of a distributed storage system is also the work of \cite{gandhi_autoscale:_2012}, which is limited to key-value datastores. 
Similarly, the work of \cite{herodotou_no_2011} is limited to Hadoop clusters.
The work of \cite{lim_automated_2010} presents policies for elastically scaling a Hadoop  storage-tier of multi-tier Web services based on automated control.
This approach is reactive and has a limited focus on Hadoop clusters.
Other examples of rule-based techniques that trigger elasticity actions are described in \cite{das2011Albatross,Bairavasundaram:2012Dynamite,han2012Lightweight}. Orthogonally to the elasticity policies, the focus in \cite{chaisiri_optimization_2012} is on the cost of resource provisioning. 
There are also several proposals on performance analysis and resource allocation (e.g., \cite{iosup_performance_2011,kitsos_adapting_2012}).

Our work also relates to proposals that employ model checking for cloud solutions and runtime quantitative verification. In the work of
\cite{Perez-Calinescu2013log2cloud}, a technique to predict the minimum VM cost of cloud deployments based on existing application logs is introduced. Queuing network theory is used to derive VM resource usage profiles. With the the latter, MDP models  are instantiated and used to verify system properties.
The work of \cite{johnson-calinescu2013incremental_verification} presents an incremental model verification technique, which is applied on component-based software systems deployed on cloud infrastructures. Their technique achieves lower verification time as only the modified components are verified. Both these techniques are used to analyze the cloud-based systems and not to drive elasticity or other adaptivity decisions.
QoSMOS \cite{calinescu2011qosmos} is a framework that also utilizes PRISM to analyse and dynamically choose the appropriate configuration of service-based systems modelled as Markov chains. Finally, the work in \cite{ghezzi2013adam} introduces a model-driven framework called ADAM, where the users provide activity diagrams that are converted to MDP models, for which cumulative rewards for various quality metrics are computed.

\section{Conclusions and Future Work}
\label{sec:conclusions}

This work presented a formal, probabilistic model checking-based approach to resizing an application cluster of VMs so that elasticity decisions are amenable to quantitative analysis. We presented MDP elasticity models and associated elasticity policies that rely on the dynamic instantiation of such models. We also conducted thorough experiments using both real and synthetic datasets, and we presented  results showing that we can significantly increase user-defined utility values and decrease the frequency of user-defined threshold violations.

An important note is that in this work we have not exploited the full potential of MDP models, which we plan to do in the future. MDP models can naturally capture complementary non-deterministic aspects of elasticity in real systems, such as provision for failure to enforce an elasticity decision and support for additional forms of elasticity like vertical resizing (e.g. resizing of CPU,RAM resources).

\section*{Acknowledgments}
This research has been co-financed by
the European Union (European Social Fund - ESF) and
Greek national funds through the Operational Program ``Education and Lifelong Learning of the National Strategic
Reference Framework (NSRF) - Research Funding Program:
Thales. Investing in knowledge society through the European Social Fund."

\appendix[PRISM implementation details]
\label{ap:prism_details}
The discussion in the previous sections regarded the models mostly at the conceptual level. In order to implement the models according to the PRISM's specifications implementation further issues need to be considered, which are briefly discussed here. More specifically, in the PRISM MDP model, each state has the following labelling variables apart from $s_{[vms\_num]}$:
\begin{itemize}

\item $previous\_action$, which denotes the type of the last action. It is important to track the previous action so that we can partially guide the model checking to specific actions. For example, when the first non-deterministic action is $add$, we are only interested in $add$ and $no\_op$ actions, as discussed earlier. In this way, we avoid the investigation of decision paths like $add_1 \rightarrow remove_2 \rightarrow add_1 \rightarrow no\_op$, which are equivalent to the $no\_op$ action.

\item $stop$, which denotes the arrival to an end component. Every state of the model can be considered as an end component, as discussed earlier. If a $no\_op$ action takes place, the current state is considered as an end component.


\item $decision$, which denotes the state type. Overall, there are three distinct state types in our model, namely $decision$ states where an action is taken, $control$ states, where the rewards are updated, and finally, $accepted$ states, which correspond to end components.
\end{itemize}

For decision making, we use cumulative rewards, which are however constructed to behave as instantaneous rewards. Cumulative state rewards with no action rewards (that can cover transition costs) are not suitable to compute the state rewards as longer paths will most probably prevail (since more values will be cumulated). Instantaneous rewards were firstly considered in the model using steps. A predefined number of steps should firstly be completed before the state reward is computed. This is a common way to model instantaneous rewards. However, the maximum number of steps had to be set arbitrarily; if not set high enough, it may not allow to reach distant states. In our system, there is no need to examine paths of actions which include both additions and removals to reach a state. Additionally, when a no operation ($no\_op$) action takes place, this indicates that currently, there is no better state to visit, so there is no reason to continue traversing the current path. Combining these two observations, we compute the state rewards as instantaneous ones only when a no operation action takes place.


\bibliographystyle{IEEEtran}
\bibliography{IEEEabrv,cloud,related}

\begin{thebibliography}{10}
\providecommand{\url}[1]{#1}
\csname url@samestyle\endcsname
\providecommand{\newblock}{\relax}
\providecommand{\bibinfo}[2]{#2}
\providecommand{\BIBentrySTDinterwordspacing}{\spaceskip=0pt\relax}
\providecommand{\BIBentryALTinterwordstretchfactor}{4}
\providecommand{\BIBentryALTinterwordspacing}{\spaceskip=\fontdimen2\font plus
\BIBentryALTinterwordstretchfactor\fontdimen3\font minus
  \fontdimen4\font\relax}
\providecommand{\BIBforeignlanguage}[2]{{%
\expandafter\ifx\csname l@#1\endcsname\relax
\typeout{** WARNING: IEEEtran.bst: No hyphenation pattern has been}%
\typeout{** loaded for the language `#1'. Using the pattern for}%
\typeout{** the default language instead.}%
\else
\language=\csname l@#1\endcsname
\fi
#2}}
\providecommand{\BIBdecl}{\relax}
\BIBdecl

\bibitem{herbst2013elasticity_definition}
N.~R. Herbst, S.~Kounev, and R.~Reussner, ``Elasticity in cloud computing: What
  it is, and what it is not,'' in \emph{Proceedings of the 10th International
  Conference on Autonomic Computing (ICAC 13)}.\hskip 1em plus 0.5em minus
  0.4em\relax Berkeley, CA: USENIX, 2013, pp. 23--27.

\bibitem{tsoumakos2013automated}
D.~Tsoumakos, I.~Konstantinou, C.~Boumpouka, S.~Sioutas, and N.~Koziris,
  ``Automated, elastic resource provisioning for nosql clusters using
  tiramola,'' in \emph{Cluster, Cloud and Grid Computing (CCGrid), 2013 13th
  IEEE/ACM International Symposium on}.\hskip 1em plus 0.5em minus 0.4em\relax
  IEEE, 2013, pp. 34--41.

\bibitem{gandhi_autoscale:_2012}
A.~Gandhi, M.~Harchol-Balter, R.~Raghunathan, and M.~A. Kozuch, ``{Autoscale:
  Dynamic, Robust Capacity Management for Multi-tier Data Centers},''
  \emph{{ACM} Transactions on Computer Systems ({TOCS)}}, vol.~30, no.~4,
  p.~14, 2012.

\bibitem{shen_cloudscale:_2011}
Z.~Shen, S.~Subbiah, X.~Gu, and J.~Wilkes, ``{Cloudscale: Elastic Resource
  Scaling for Multi-tenant Cloud Systems},'' in \emph{SOCC}, 2011, pp.
  5:1--5:14.

\bibitem{gong_press:_2010}
Z.~Gong, X.~Gu, and J.~Wilkes, ``{Press: Predictive Elastic Resource Scaling
  for Cloud Systems},'' in \emph{{CNSM)}}, 2010, pp. 9--16.

\bibitem{trushkowsky_scads_2011}
B.~Trushkowsky, P.~Bodík, A.~Fox, M.~J. Franklin, M.~I. Jordan, and D.~A.
  Patterson, ``The {SCADS} director: Scaling a distributed storage system under
  stringent performance requirements.'' in \emph{{FAST}}, 2011, pp. 163--176.

\bibitem{Bairavasundaram:2012Dynamite}
L.~N. Bairavasundaram, G.~Soundararajan, V.~Mathur, K.~Voruganti, and
  K.~Srinivasan, ``Responding rapidly to service level violations using virtual
  appliances,'' \emph{SIGOPS Oper. Syst. Rev.}, vol.~46, no.~3, pp. 32--40,
  2012.

\bibitem{KC03}
J.~O. Kephart and D.~M. Chess, ``The vision of autonomic computing.''
  \emph{IEEE Computer}, vol.~36, no.~1, pp. 41--50, 2003.

\bibitem{CGK+12}
R.~Calinescu, C.~Ghezzi, M.~Z. Kwiatkowska, and R.~Mirandola, ``Self-adaptive
  software needs quantitative verification at runtime,'' \emph{Commun. ACM},
  vol.~55, no.~9, pp. 69--77, 2012.

\bibitem{FKNP11}
V.~Forejt, M.~Kwiatkowska, G.~Norman, and D.~Parker, ``Automated verification
  techniques for probabilistic systems,'' in \emph{Formal Methods for Eternal
  Networked Software Systems (SFM'11)}, ser. LNCS, vol. 6659, 2011, pp.
  53--113.

\bibitem{kwiatkowska2009Prism}
M.~Kwiatkowska, G.~Norman, and D.~Parker, ``Prism: probabilistic model checking
  for performance and reliability analysis,'' \emph{SIGMETRICS Perform. Eval.
  Rev.}, vol.~36, no.~4, pp. 40--45, 2009.

\bibitem{puterman1994MDP}
M.~L. Puterman, \emph{Markov Decision Processes: Discrete Stochastic Dynamic
  Programming}, 1st~ed.\hskip 1em plus 0.5em minus 0.4em\relax New York, NY,
  USA: John Wiley \& Sons, Inc., 1994.

\bibitem{qureshi2009cutting}
A.~Qureshi, R.~Weber, H.~Balakrishnan, J.~Guttag, and B.~Maggs, ``Cutting the
  electric bill for internet-scale systems,'' \emph{ACM SIGCOMM Computer
  Communication Review}, vol.~39, no.~4, pp. 123--134, 2009.

\bibitem{chen2008energy}
G.~Chen, W.~He, J.~Liu, S.~Nath, L.~Rigas, L.~Xiao, and F.~Zhao, ``Energy-aware
  server provisioning and load dispatching for connection-intensive internet
  services.'' in \emph{NSDI}, vol.~8, 2008, pp. 337--350.

\bibitem{KoukisVK13}
V.~Koukis, C.~Venetsanopoulos, and N.~Koziris, ``$\sim$okeanos: Building a
  cloud, cluster by cluster,'' \emph{IEEE Internet Computing}, vol.~17, no.~3,
  pp. 67--71, 2013.

\bibitem{massie2004ganglia}
M.~L. Massie, B.~N. Chun, and D.~E. Culler, ``The ganglia distributed
  monitoring system: design, implementation, and experience,'' \emph{Parallel
  Computing}, vol.~30, no.~7, pp. 817--840, 2004.

\bibitem{marshall_elastic_2010}
P.~Marshall, K.~Keahey, and T.~Freeman, ``{Elastic Site: Using Clouds to
  Elastically Extend Site Resources},'' in \emph{{CCGID}}, 2010, pp. 43--52.

\bibitem{elmroth_self-management_2011}
E.~Elmroth, J.~Tordsson, F.~Hernández, A.~Ali-Eldin, P.~Svärd, M.~Sedaghat,
  and W.~Li, ``{Self-management Challenges for Multi-Cloud Architectures},'' in
  \emph{Towards a Service-Based Internet}.\hskip 1em plus 0.5em minus
  0.4em\relax Springer, 2011, pp. 38--49.

\bibitem{wang_flexible_2011}
C.~Wang, K.~Schwan, V.~Talwar, G.~Eisenhauer, L.~Hu, and M.~Wolf, ``{A Flexible
  Architecture Integrating Monitoring and Analytics for Managing Large-scale
  Data Centers},'' in \emph{{ICAC}}, 2011, pp. 141–--150.

\bibitem{herodotou_no_2011}
H.~Herodotou, F.~Dong, and S.~Babu, ``{No One (Cluster) Size Fits All:
  Automatic Cluster Sizing for Data-intensive Analytics},'' in \emph{{SOCC}},
  2011, p.~18.

\bibitem{lim_automated_2010}
H.~C. Lim, S.~Babu, and J.~S. Chase, ``{Automated Control for Elastic
  Storage},'' in \emph{{ICAC}}, 2010, pp. 1--10.

\bibitem{das2011Albatross}
S.~Das, S.~Nishimura, D.~Agrawal, and A.~El~Abbadi, ``Albatross: lightweight
  elasticity in shared storage databases for the cloud using live data
  migration,'' \emph{Proceedings of the VLDB Endowment}, vol.~4, no.~8, pp.
  494--505, 2011.

\bibitem{han2012Lightweight}
R.~Han, L.~Guo, M.~M. Ghanem, and Y.~Guo, ``Lightweight resource scaling for
  cloud applications,'' in \emph{Cluster, Cloud and Grid Computing (CCGrid),
  2012 12th IEEE/ACM International Symposium on}.\hskip 1em plus 0.5em minus
  0.4em\relax IEEE, 2012, pp. 644--651.

\bibitem{chaisiri_optimization_2012}
S.~Chaisiri, B.-S. Lee, and D.~Niyato, ``{Optimization of Resource Provisioning
  Cost in Cloud Computing},'' \emph{Services Computing, {IEEE} Transactions
  on}, vol.~5, no.~2, pp. 164–--177, 2012.

\bibitem{iosup_performance_2011}
A.~Iosup, S.~Ostermann, M.~N. Yigitbasi, R.~Prodan, T.~Fahringer, and D.~H.
  Epema, ``{Performance analysis of cloud computing services for many-tasks
  scientific computing},'' \emph{Parallel and Distributed Systems, {IEEE}
  Transactions on}, vol.~22, no.~6, pp. 931–--945, 2011.

\bibitem{kitsos_adapting_2012}
I.~Kitsos, A.~Papaioannou, N.~Tsikoudis, and K.~Magoutis, ``{Adapting
  Data-intensive Workloads to Generic Allocation Policies in Cloud
  Infrastructures},'' in \emph{{NOMS}}, 2012, pp. 25–--33.

\bibitem{Perez-Calinescu2013log2cloud}
D.~Perez-Palacin, R.~Calinescu, and J.~Merseguer, ``Log2cloud: Log-based
  prediction of cost-performance trade-offs for cloud deployments,'' in
  \emph{Proceedings of the 28th Annual ACM Symposium on Applied Computing},
  2013, pp. 397--404.

\bibitem{johnson-calinescu2013incremental_verification}
K.~Johnson, R.~Calinescu, and S.~Kikuchi, ``An incremental verification
  framework for component-based software systems,'' in \emph{Proceedings of the
  16th International ACM Sigsoft Symposium on Component-based Software
  Engineering}, ser. CBSE '13, pp. 33--42.

\bibitem{calinescu2011qosmos}
R.~Calinescu, L.~Grunske, M.~Kwiatkowska, R.~Mirandola, and G.~Tamburrelli,
  ``Dynamic qos management and optimization in service-based systems,''
  \emph{Software Engineering, IEEE Transactions on}, vol.~37, no.~3, pp.
  387--409, 2011.

\bibitem{ghezzi2013adam}
C.~Ghezzi, L.~S. Pinto, P.~Spoletini, and G.~Tamburrelli, ``Managing
  non-functional uncertainty via model-driven adaptivity,'' in
  \emph{Proceedings of the 2013 International Conference on Software
  Engineering}.\hskip 1em plus 0.5em minus 0.4em\relax IEEE Press, 2013, pp.
  33--42.

\end{thebibliography}





\end{document}